\pdfoutput=1

\documentclass[11pt]{article}

\usepackage[final]{exploracoder}

\usepackage{times}
\usepackage{latexsym}

\usepackage[T1]{fontenc}

\usepackage[utf8]{inputenc}

\usepackage{microtype}

\usepackage{inconsolata}

\usepackage{graphicx}

\usepackage{comment}
\usepackage{hyperref}
\usepackage{booktabs}
\usepackage{multirow}
\usepackage{hhline}
\usepackage{colortbl}
\usepackage{listings}
\usepackage{subcaption} 
\usepackage{enumitem}
\usepackage{amsmath} 
\usepackage{pifont} 

\usepackage{caption}     
\usepackage{stfloats}    
\usepackage{listings}    
\usepackage{pdflscape}   
\usepackage{longtable}   
\newcommand{\yes}{\textcolor{green}{\ding{51}}} 
\newcommand{\no}{\textcolor{red}{\ding{55}}}     

\lstdefinestyle{mystyle}{
    basicstyle=\ttfamily\tiny,
    breakatwhitespace=true,
    breaklines=true,
    keepspaces=true,
    showspaces=false,
    showstringspaces=false,
    frame=single,
    extendedchars=false,
    inputencoding=utf8
}
\title{ExploraCoder: Advancing Code Generation for Multiple Unseen APIs\\ via Planning and Chained Exploration}

\author{
 \textbf{Yunkun Wang\textsuperscript{1}\thanks{Work is done during internship at Alibaba Group.}},~~
 \textbf{Yue Zhang\textsuperscript{2}},~~
 \textbf{Zhen Qin\textsuperscript{1}},~~
 \textbf{Chen Zhi\textsuperscript{1}\footnotemark[2]},
\\
 \textbf{Binhua Li\textsuperscript{2}},~~
 \textbf{Fei Huang\textsuperscript{2}},~~
 \textbf{Yongbin Li\textsuperscript{2}\thanks{Corresponding authors.}},~~
 \textbf{Shuiguang Deng \textsuperscript{1}\footnotemark[2]}
\\
 \textsuperscript{1}Zhejiang University,~~
 \textsuperscript{2}Alibaba Group,
\\
\texttt{\{wangykun, zhenqin, zjuzhichen, dengsg\}@zju.edu.cn}\\
\texttt{\{shiyu.zy, binhua.lbh, f.huang, shuide.lyb\}@alibaba-inc.com}\\
}

\begin{document}
\maketitle
\begin{abstract}

Large language models face intrinsic limitations in coding with APIs that are unseen in their training corpora. As libraries continuously evolve, it becomes impractical to exhaustively retrain LLMs with new API knowledge. This limitation hampers LLMs from solving programming problems which require newly introduced or privately maintained libraries. 
Inspired by exploratory programming paradigm in human behavior, we propose \textbf{ExploraCoder}, a training-free framework that empowers LLMs to invoke multiple unseen APIs in code solution by (1) planning a complex problem into several API invocation subtasks, and (2) experimenting with correct API usage at intermediate steps through a novel chain-of-API-exploration.
We conduct evaluation on program synthesizing tasks involving complex API interactions. Experimental results demonstrate that ExploraCoder significantly improves performance for models lacking prior API knowledge, achieving absolute increases of up to 11.99\% over retrieval-based approaches and 17.28\% over pretraining-based methods in pass@10.

\end{abstract}

\section{Introduction}
Library-oriented code generation refers to the automatic generation of code that utilizes specified library's APIs to solve programming problems \citep{zan_cert_2022,liu2023codegen4libs}. 
This task becomes particularly complex when the solution requires the integration of multiple APIs from the library, demanding not only knowledge of individual API functionalities but also an understanding of their interactions and dependencies \citep{alrubaye2019learningrecommendthirdpartylibrary, zan_diffcoder_nodate}.
Modern large language model (LLM), such as ChatGPT \citep{chatgpt} and CodeLlaMA 
\citep{rozière2024codellamaopenfoundation}, has demonstrated remarkable capability in generating API invocations using prior knowledge from pretraining stage \citep{zan_private-library-oriented_2023}.
However, a significant challenge arises when the target API knowledge is sparse, outdated, or entirely unseen in the training data. This limitation hampers LLMs from problem solving that requires newly introduced or privately maintained libraries.

Prior work proposed to use continual pretraining \citep{gururangan2020dontstoppretraining} to address this knowledge gap \citep{zan_cert_2022}. But this is often impractical due to the scarcity of training data for new libraries and the substantial costs of retraining LLMs.
Another line of work adopts a standard retrieval-augmented generation (RAG) framework for unseen API invocations \citep{zhou_docprompting_2023,zan_private-library-oriented_2023,liu2023codegen4libs}, where LLM acquires API knowledge from retrieving the library documentation. While effective for simple API invocation tasks, these methods struggle with complex scenarios requiring multiple API invocations \citep{zan_private-library-oriented_2023,zan_diffcoder_nodate,ma_compositional_2024}.

More recent studies propose to address complex API invocation tasks by improving document-retrieval \citep{ma_compositional_2024} and preactively planning coding steps \citep{li2024epigen}. However, they overlook the challenge posed by the potential ambiguities in the API documentation. Some work adopts iterative or agentic workflow \citep{olausson_is_2024,yao2022react,zhu2024knowagent} to reactively plan for retrieval and debugging, however, the end-to-end code construction could still expose the limitations of LLMs in coordinating multi-API interactions.

\begin{figure*}[t]
\centering
\includegraphics[width=1\textwidth]{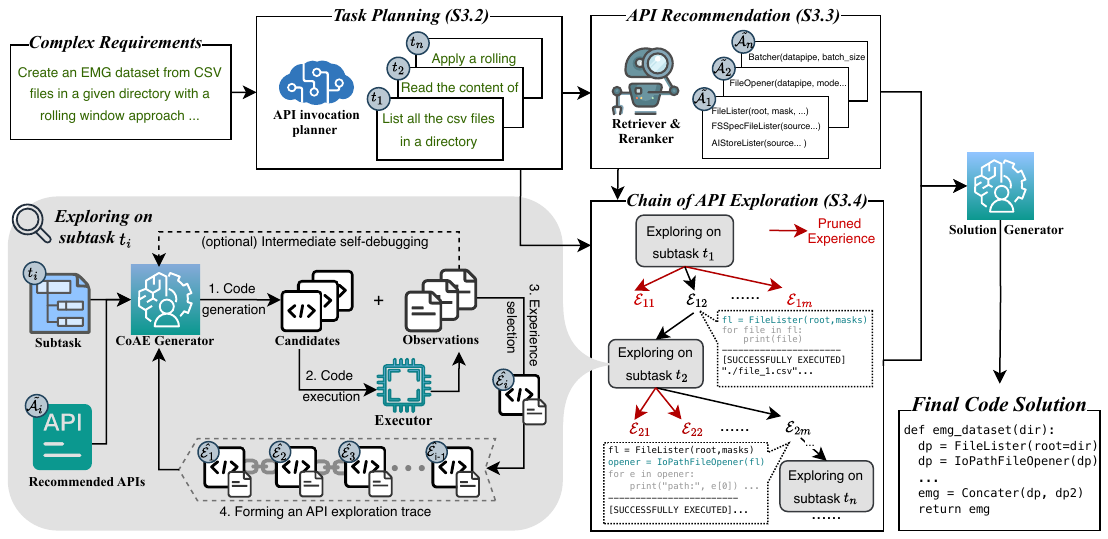}
\caption{An Overview of ExploraCoder Framework. ExploraCoder processes the given problem through \emph{\textbf{Task Planning}}, \emph{\textbf{API Recommendation}}, and \emph{\textbf{Chain of API Exploration}} modules. The \setlength{\fboxsep}{0.3pt}{\colorbox{gray!20}{\strut gray block}} in the bottom-left corner illustrates the detailed inner process in the Chain of API Exploration. Finally, the processed results are used by a solution generator to generate final code solutions for the programming problem.}
\label{fig:exploracoder_framework}
\vspace{-5mm}
\end{figure*}

When coding with an unfamiliar library, experienced developers would adpot an \emph{Exploratory Programming paradigm} \citep{1986exploratory,exploratory_programming}. This involves first understanding the library’s capabilities through documentation to devise a broad plan, and then actively experimenting with individual API calls to gain practical experience, ultimately leading to a correct code solution. Inspired by this behavior, we propose \textbf{ExploraCoder}, a training-free framework aiming to facilitate LLM to invoke multiple unseen APIs. 
As shown in Figure \ref{fig:exploracoder_framework}, given a complex programming problem, ExploraCoder begins by planning a series of simpler API invocation subtasks based on library documentation. For each subtask, it recommends a set of candidate APIs. Subsequently, a Chain of API Exploration (CoAE) is performed, iteratively experimenting with various subtask-wise API invocations while passing valuable usage insights to the plannings of subsequent subtasks. This process forms an API exploration trace, which facilitates the LLM in deriving the final solution.

Evaluating unseen-library-oriented code generation requires an unexposed library. Prior works \citep{zan_private-library-oriented_2023,ma_compositional_2024} have created simple benchmarks using the Torchdata library, as it strikes a balance between minimum client code use for researcher's problem designing and limited exposure to modern LLMs.
However, benchmarking multiple unseen-API tasks remains underexplored.
To address this gap and better reflect real-world complex programming challenges, we constructed a new Torchdata-based benchmark, \textbf{Torchdata-Manual}, featuring complex multi-API problems.
Experimental results on our Torchdata-Manual and an existing Torchdata-Github benchmarks demonstrate that ExploraCoder significantly improves performance for models lacking prior API knowledge, achieving absolute gains of up to 11.99\% over various retrieval-based approaches and 17.28\% over pretraining methods in pass@10. Moreover, we find the integration of an intermediate self-debug mechanism further boosts ExploraCoder’s performance on more challenging tasks. 

This paper makes the following contributions:
\begin{itemize}[leftmargin=0.4cm]

\item We propose ExploraCoder, a unified framework that incorporates unseen API knowledge from documentation into a novel step-wise code generation method, Chain-of-API-Exploration. By leveraging this framework, LLMs can plan based on library documentation and actively experiment with APIs in intermediate steps, mirroring the Exploratory Programming paradigm employed by human developers.

\item We construct Torchdata-Manual, a new library-oriented benchmark that, to the best of our knowledge, features the highest number of API invocations per task among publicly reported executable library-oriented benchmarks. The code and data are available at~ \url{https://github.com/greenlight2000/ExploraCoder}.

\item Experimental results and case studies on ours and an existing benchmark demonstrate ExploraCoder’s superior performance on multi-API tasks compared to competitive baselines.

\end{itemize}

\section{Related Work}

\paragraph{Complex Code Generation with LLM.}
Code generation, the process of producing code from NL specifications, has seen remarkable advancements with LLMs \citep{openai2024gpt4technicalreport, yan2024codescopeexecutionbasedmultilingualmultitask}. Recent research has increasingly focused on LLMs tackling complex coding tasks, such as competition \citep{Li_2022codeforeces}, library-oriented \citep{bogomolov2024longcode}, and repo-level \citep{jimenez2024swebench} problems. One prominent paradigm leverages chain-of-thought \citep{wei_cot_2022} to plan for intermediate steps before complex code generation, whereas its effectiveness diminishes when high quality plans cannot be derived \citep{jiang_self-planning_2024}.
Another direction proposes to debug after the generation of code \citep{olausson_is_2024}, but they typically require accessibility of test cases. 
ExploraCoder distinguishes itself by applying an intermediate code construction that leverages executability signals to rectify coding plans at steps in real time.

\paragraph{Library-Oriented Code Generation.}
Real-world programming problems often involve the use of external libraries, posing a challenge for LLMs to invoke APIs unseen from training data. 
Continued pretraining on the new API data, though intuitive, is often impractical due to its complexity and cost. Most prior studies adopt a naive RAG framework as an alternative to incorporate APIs knowledge from library documentation \citep{zhou_docprompting_2023, zan_private-library-oriented_2023}. But they struggle with more complex problems that require multiple API invocations \citep{zan_diffcoder_nodate}. Recent studies have attempted to improve the RAG frameworks. For examples, CAPIR \citep{ma_compositional_2024} proposed a decomposed the retrieval process to identify accurate API docs. EpiGen \citep{li2024epigen} makes preactive NL plans for one-pass code generation. These works mainly focus on preprocessing relevant API context, while overlooking the reasoning limitation of LLMs in multi-API interactions, and the challenge posed by the potential ambiguity in API documentation.

\paragraph{Unseen Library Benchmarks.}
Constructing unseen library benchmarks is particularly challenging, as libraries new enough to have limited exposure to modern LLMs often lack the rich client code needed for developing complex problems.
Previous work has generally turned to a Torchdata library to manually build small-scale API invocation benchmarks. 

\citet{zan_private-library-oriented_2023} constructed TorchdataEval mostly involving 1-2 simple API invocations. \citet{ma_compositional_2024} introduced 50 multi-API programming tasks adapted from Torchdata client code from Github, each involving 3–8 API invocations. However, some of its tasks remain relatively simple and real-world development often involves more API interactions \citep{api_client_use,method_call_use}. This gap highlights the need for a more complex unseen library benchmark.

\section{ExploraCoder Framework}
\subsection{Task Definition}

This work addresses the task of library-oriented code generation \citep{zan_cert_2022}. Formally, given a problem \(\psi\) that specifies the user requirement and a library API documentation $\mathcal{A}$, a model $\theta$ generates code solutions $p\sim \mathcal{P}_\theta(.|\psi,A)$.

Most code libraries provide basic information about their APIs, such as API signatures, descriptions, and high-level library overviews with minimum example usage code. In this paper, we assume the accessibility of this information from API documents. As shown in Figure \ref{fig:exploracoder_framework}, ExploraCoder will automatically identify relevant subset of APIs $\mathcal{\hat A}$ and accumulate useful experience of intermediate API invocations $\hat{\mathcal{E}}$, both of which are then used as augmenting signals to generate the final solution:
\begin{equation}
p := ExploraCoder(\psi, \hat{\mathcal{A}}, \hat{\mathcal{E}})
\end{equation}

\subsection{Planning for API invocation}
\label{sec: task_planning_module}
Real-world programming problems often involve composite operations \citep{yu2024codereval}, necessitating a plan for where and how APIs can contribute to problem-solving. Specifically, we need to outline several API-related subtasks, upon which ExploraCoder will sequentially explore the correct API calls. Ideally, we aim to set the planning granularity to simple subtasks where each requires only 1–2 unseen API invocations.
However, the functional granularity of APIs is domain-specific, often falling out of distribution (OOD) of LLMs when the library is absent from their training data, posing a challenge in aligning task planning with typical API usage patterns.

To address this, we leverage the in-context learning capabilities of LLMs \citep{an2023context} by providing a condensed library overview and a small number of planner examples. 
This enables the LLMs to learn high-level usage patterns of the library without needing to know all its APIs. In this work, we prompt GPT-3.5-turbo-0125 to automatically summarize a piece of text $s$ from the library overview and extract few-shot planners $\mathcal{D} = \{ \langle \psi_j, \{ t_u \}_{u=1}^{w_j} \rangle\}^{n_\mathcal{D}}_{j=1}$ from the provided code examples, where $\psi_j$ is the requirement of the $j$-th code example, and $t_u$ is the explanation of $u$-th API invocation. Note that we do not leak any detailed API usage or benchmark-related knowledge to models (detailed in Appendix \ref{app: implementation_details}). Now, we can plan $n$ API-related subtasks for a given problem $\psi$:
\begin{equation}
\{t_i\}^n_{i=1}\sim \mathcal{P}_{\theta}(.|\psi, \mathcal{D}, s)
\end{equation}

\subsection{API Recommendation}
The API recommendation module serves to recommend relevant API documents $\mathcal{A}_i = \{ a^{(1)}, \dots, a^{(k)} \}$ for each API invocation subtask  $t_i$ . We process the documents into tabular retrieval pool, where each row consists of the API import path, signature, and description. 
We first use a dense retriever to retrieve an initial set of APIs by computing the similarity between  $t_i$ and each $a_j$. 
\begin{equation}
{\mathcal{A}}_i = \text{top-}k\ \{ \text{sim}(a_j, t_i) \mid a_j \in \mathcal{A} \}
\end{equation}
Then, we prompt LLM to re-rank and drop irrelevant APIs for each subtask, providing a refined subset $\{\mathcal{\tilde{A}}_i\}^n_{i=1}$ for Chain of API Exploration, where then the actually used APIs $\tilde{\mathcal{A}}_\text{CoAE}$ will be recorded. Meanwhile, we also conduct an inter-task reranking \citep{ma_compositional_2024} to recommend $k_G$ APIs $\tilde{\mathcal{A}}_G$ from a global perspective. In the final solution stage, we provide for the generator:
\begin{equation}
\hat{\mathcal{A}} = \tilde{\mathcal{A}}_{\text{CoAE}} \cup \tilde{\mathcal{A}}_{G}
\end{equation}

\subsection{Chain of API Exploration}

Previous work shows LLMs struggle to directly invoke multiple unseen APIs in a single run \citep{zan_diffcoder_nodate}. The challenge arises from LLMs' tendency to hallucinate unfamiliar APIs usage. Hallucinations in early decoding step could compromise subsequent API calls due to the autoregressive nature of LLM, further compounding the error.

In contrast, when lacking knowledge of relevant APIs, developers could adopt an exploratory programming paradigm, actively experimenting with partial code in a sandbox environment to accumulate correct API usage experience. Inspired by this behavior, we designed a Chain of API Exploration (CoAE) to sequentially explore API usage and solve the $n$ subtasks $\{ \langle \tilde{\mathcal{A}}_i, t_i \rangle \}_{i=1}^n$. We now formalize the main steps in CoAE.

\paragraph{Experimental code generation.}
We prompt the LLM to generate $m$ diversified experimental code snippets for intermediate subtask \(t_i\):
\begin{equation}
\{p_{i,j}\}^m_{p_{j=1}} \sim \mathcal{P}_\theta(.|t_i, s, \mathcal{\hat{A}}_i,\mathcal{E}_{1:i-1})
\end{equation}
where $s$ is the high-level library information from Section \ref{sec: task_planning_module}, and $\mathcal{E}_{1:i-1}$ is the accumulated invocation experience from prior subtasks that could further enhance the preactive planning of $t_i$. 
We define API invocation experience as the combination of an intermediate code snippet and its execution output, which is elaborated in the next paragraph. Each experimental code will attempt to solve the subtask by making different API invocations, and print out valuable usage knowledge. Such feedback will be observed by LLM in the next step.

\paragraph{Code execution and observation.}
At each subtask, LLM is encouraged to print out insightful information to expand API usage knowledge, such as format of the current API returned object that could be used as input in other subtasks.
We capture the output from directly executing the experimental code in a sandbox environment.
Specifically, given $t_i$ and $p_{i,j}$, the observation $o_{i,j}$ by the LLM consists of the codes’ executability $\delta$ , error message $\varepsilon$, and program output $\gamma$. We now can assemble $m$ candidate API invocation experience for $t_i$ as:
\begin{equation}
\mathcal{E}_{i} = \{\langle t_i, p_{i,j},o_{i,j} \rangle\}^m_{j=1}
\end{equation}

\paragraph{Experience exploitation by intermediate self-debugging.} 
In our preliminary experiments, we found experimental codes often fail to execute due to simple mistakes (e.g., missing import statements). Additionally, some challenging subtasks require complex API interactions with prior subtasks, which LLMs struggle to solve. This hinders the acquisition of additional API usage insights, and the intermediate failures could potentially degenerate the performance of exploration chain.
To address this, we prompt the LLM to debug the codes when all candidate codes for a given subtask fail to execute, thereby enhancing its usage experience. We report the effectiveness of ExploraCoder, both with and without debugging mechanism. 

\paragraph{Experience selection strategy.}
After obtaining $m$ candidate exploration experience  $\{\mathcal{E}_{i,j}\}^m_{j=1}$  on $t_i$ . The goal in this step is to select the most valuable one $\mathcal{\hat{E}}_i$ and prune the others for $t_i$ .
In this work, we adopt a simple but effective selection strategy: (1) randomly select a candidate that has successfully executed, prioritizing the ones with valid outputs; (2) if all candidates fail to execute, we randomly select a failed one. Then, the selected experience will be passed on to the next subtask and accumulates progressively. Ultimately, we obtain an API exploration trace of the following form to aid in solution generation:
\begin{equation}
\mathcal{\hat{E}} = \{\mathcal{\hat{E}}_i\}^n_{i=1}
\end{equation}

\section{Benchmark Construction}
Unseen library benchmarks are essential for evaluating retrieval-based methods in handling unseen APIs. Existing benchmarks typically involve simple API invocations or apply lexical-based evaluation metrics. To provide rigorous evaluation of complex unseen API invocations, we aim to construct \textbf{execution-based multi-API} benchmark that remain \textbf{untrained} on representative LLMs. Following prior work \citep{zan_diffcoder_nodate}, we use Torchdata-based evaluation, which remains unexposed to powerful LLMs such as GPT-3.5 and GPT-4-0613, while allowing knowledge acquisition by newer models. This provides a valuable reference point for assessing approaches across LLMs with varying levels of API prior knowledge.

\paragraph{Torchdata-Manual.} We developed a new benchmark called Torchdata-Manual, comprising 100 manually crafted programming problems. Each problem involves 8-14 distinct Torchdata APIs. To ensure the diversity of the programming tasks, we randomly sampled numerous API combinations from the Torchdata documentation and selected plausible combinations to formulate the problem. Two programmers with more than five years of Python coding experience are invited to review the benchmark. More detailed construction methodology is provided in the Appendix \ref{app: TorchdataManual_construction}. To the best of our knowledge, Torchdata-Manual features the longest API sequences among publicly reported execution-based library-oriented benchmarks.

\paragraph{Torchdata-Github.} We also evaluate on an existing benchmark \citep{ma_compositional_2024}, including 50 Torchdata problems adapted from client project of Torchdata on GitHub, featuring coarse-grained user requirements that entails 3-8 API invocations. We curated the dataset by manualy supplementing external resources needed to run test cases in some problems\footnote{Some external resources, such as local files to be loaded in problems, are not provided by \citet{ma_compositional_2024}.} and named it as Torchdata-Github.

\paragraph{MonkBeatEval.} To test generalizability beyond Torchdata-based evaluation, we also adapted an existing multi-library benchmark for unseen settings, with results reported in Appendix \ref{app:monkbeateval-baselines}.

\section{Experiments}
\label{sec: experiments}
\subsection{Experimental setups}
\paragraph{Benchmarks and base language models.}
We evaluate ExploraCoder on Torchdata-Github and Torchdata-Manual benchmarks.
Based on the the publicly available information on models’ training data cutoff date, we conduct our main experiments under two base models settings: (1) \emph{API-untrained model}, where the API knowledge is unseen by model during training phase. We choose GPT-3.5-turbo-0125 and GPT-4-0613 as representatives.
(2) \emph{API-pretrained model}, where the API knowledge is pretrained in model. We represent it by GPT-4-1106-preview and two SOTA opensource code LLM: CodeQwen-1.5 and DeepseekCoder-6.7b.
Due to the token budgets, we primarily experiment ExploraCoder with GPT-3.5-turbo-0125, while reporting GPT-4-0613 results where necessary to further support our conclusions.

\paragraph{Evaluation metrics.}
We adopt \textbf{Pass@k} as our primary evaluation metrics.
For each problem, we randomly sample $n\ge k$ code solutions from the model to execute against test cases. And pass@k is calculated as the percentage of problems solved using k candidates.
To better observe nuance differences in harder problems, we additionally report \textbf{Success@k} \citep{chen2024codemetrics} which relaxes the evaluation criteria by measuring whether the generated code can be executed successfully without runtime errors within limited timeout constraints.

\paragraph{Implementation details.} We implement ExplorCoder by setting $k_\mathcal{D}=4$ for task planning. For API recommendation, we set $k=20$ as initial retrieval volume, $k_{G}=15$ on Torchdata-Github following \citet{ma_compositional_2024} and $k_{G}=20$ on Torchdata-Manual. For CoAE, we set $m=5$. To generate diverse candidates, we set the $temperature=0.8$ and $top\_p=0.95$ for our CoAE and final solution generation across all baselines. More detailed experimental settings are left in Appendix \ref{app: implementation_details}

\begin{table*}
    \centering
    
    \resizebox{\textwidth}{!}{
    \begin{tabular}{clcccccccc}
    \hline
    
    && \multicolumn{2}{c}{\bfseries{$k=1$}} & \multicolumn{2}{c}{\bfseries{$k=5$}} & \multicolumn{2}{c}{\bfseries{$k=10$}} & \multicolumn{2}{c}{\bfseries{$k=20$}} \\
    \cmidrule(lr){3-4} \cmidrule(lr){5-6} \cmidrule(lr){7-8} \cmidrule(lr){9-10}
    API Knowledge&\multicolumn{1}{c}{\centering Method}& \multicolumn{1}{c}{Pass}& Success& Pass& Success& Pass& Success& Pass& Success\\
    \hline
    \multicolumn{2}{c}{\cellcolor{gray!20} } & \multicolumn{8}{c}{\cellcolor{gray!20} Torchdata-Github} \\
    \multirow{3}{*}{\shortstack{Pretrained\\in models}}
    &DeepSeekCoder-6.7B& \multicolumn{1}{|c}{5.24\%}& 6.86\%& 14.43\%& 19.28\% & 18.64\%& 27.38\%& 21.80\%& 37.23\%\\
    &CodeQwen1.5-7B& \multicolumn{1}{|c}{3.24\%}& 6.10\%& 11.60\%& 19.94\%& 16.57\%& 28.56\%& 19.90\%& 37.42\%\\
    &GPT-4-1106-preview&  \multicolumn{1}{|c}{7.43\%}& 11.52\%& 16.19\%& 28.88\%& 21.34\%&  38.74\%&  25.81\%&  45.71\%\\
    \hline

    \multirow{6}{*}{\shortstack{Untrained\\in models}}
    &GPT-3.5-turbo-0125& \multicolumn{1}{|c}{1.70\%}& 2.09\%& 5.54\%& 6.95\%& 7.28\%& 9.64\%& 8.00\%& 11.90\%\\
    & + naive RAG & \multicolumn{1}{|c}{6.00\%}& 10.57\%& 10.55\%& 24.00\%& 14.67\%& 32.50\%& 20.83\%& 40.81\%\\
    & \textbf{+ ExploraCoder}& \multicolumn{1}{|c}{\textbf{10.19\%}}& \textbf{19.50\%}& \textbf{18.64\%}& \textbf{39.39\%}& \textbf{21.67\%}&\textbf{48.56\%}&\textbf{25.62\%}& \textbf{57.30\%} \\
    \cline{2-10}
    & GPT-4-0613 & \multicolumn{1}{|c}{3.50\%}& 5.43\%& 8.86\%& 16.35\%& 11.45\%& 23.79\%& 13.80\%& 31.52\%\\
    & + naive RAG & \multicolumn{1}{|c}{10.09\%}& \textbf{29.64\%}& 20.11\%& 39.04\%& 24.07\%& 45.16\%& 27.81\%& 49.33\%\\
    & \textbf{+ ExploraCoder}& \multicolumn{1}{|c}{\textbf{15.43\%}}& 23.10\%& \textbf{21.53\%}& \textbf{45.62\%}& \textbf{28.11\%}& \textbf{55.25\%}& \textbf{30.00\%}&\textbf{61.87\%} \\
    \hline
    \hline
    \multicolumn{2}{c}{\cellcolor{gray!20} } & \multicolumn{8}{c}{\cellcolor{gray!20}Torchdata-Manual} \\
    \multirow{5}{*}{\shortstack{Pretrained\\in models}} &DeepSeekCoder-6.7B& \multicolumn{1}{|c}{0\%}& 0.48\%& 0\%& 1.57\%& 0\%& 1.95\%& 0\%& 2.00\%\\
    &CodeQwen1.5-7B& \multicolumn{1}{|c}{0\%}& 0.39\%& 0\%& 1.43\%& 0\%& 2.86\%& 0\%& 5.71\%\\
    &   GPT-4-1106-preview&  \multicolumn{1}{|c}{0.16\%}& 1.37\%& 0.71\%& 6.28\%& 1.62\%&  11.56\%&  2.79\%&  20.89\%\\
    & + naive RAG& \multicolumn{1}{|c}{3.19\%}& 6.38\%& 12.15\%& 22.15\%& 18.30\%& 31.46\%& 24.11\%&39.11\%\\
    & + ExploraCoder & \multicolumn{1}{|c}{14.62\%} & 32.77\%& 31.19\%& 57.03\%& 37.56\%& 63.47\%& 42.20\%& 67.73\%\\
    \hline

    \multirow{6}{*}{\shortstack{Untrained\\in models}}
    &GPT-3.5-turbo-0125& \multicolumn{1}{|c}{0\%}& 0\%& 0\%& 0\%& 0\%& 0\%& 0\%& 0\%\\
    & + naive RAG & \multicolumn{1}{|c}{0.19\%}& 0.615\%& 0.89\%& 2.92\%& 1.66\%& 5.475\%& 2.81\%& 9.53\%\\
    & \textbf{+ ExploraCoder}& \multicolumn{1}{|c}{\textbf{7.00\%}} & \textbf{14.80\%}& \textbf{11.54\%}& \textbf{22.89\%}& \textbf{13.84\%}&\textbf{25.40\%}&\textbf{15.67\%} & \textbf{27.56\%} \\
    \cline{2-10}
    & GPT-4-0613 & \multicolumn{1}{|c}{0\%}& 0.05\%& 0\%& 0.23\%& 0\%& 0.465& 0\%& 0.93\%\\
    & + naive RAG & \multicolumn{1}{|c}{1.12\%}& 2.94\%& 3.37\%& 8.66\%& 4.68\%& 11.98\%& 6.67\%& 16.36\%\\
    & \textbf{+ ExploraCoder}& \multicolumn{1}{|c}{\textbf{16.49\%}}& \textbf{24.16\%}& \textbf{26.10\%}& \textbf{36.89\%}& \textbf{29.41\%}& \textbf{40.68\%}& \textbf{33.32\%}&\textbf{44.32\%}\\
    \hline
    
    \end{tabular}
    }
    \caption{Evaluation of LLMs with varying levels of prior API knowledge. We apply document-retrieval to augment the API-untrained models across two datasets, and the underperforming API-pretrained GPT4 on Torchdata-Manual.}
    \label{tab:Torchdata-Github-Main}
    \vspace{-15pt}
\end{table*}
\subsection{Multi-API invocations using LLMs with varying prior API knowledge}
\label{subsec: exp_RQ1}

We consider pretraining and document-retrieval as two API knowledge integration paradigms, and analyze their effectiveness in complex multi-API generation task in Table \ref{tab:Torchdata-Github-Main}.

\paragraph{Invoking APIs using API-untrained and API-pretrained models.}

By analyzing the direct generation performance of the five base models, we observe that API-pretrained models consistently outperform API-untrained models. This highlights the importance of prior API knowledge in library-oriented code generation. And the lower performance across all models on the Torchdata-Manual further underscores the challenge posed by more complex API invocations, making it a more effective benchmark for evaluation.

Through a naive RAG framework \citep{zhou_docprompting_2023}, the performance of API-untrained models has been effectively improved, bridging the gap caused by the lack of prior API knowledge. 
We make an indirect comparison of retrieval and pretraining methods by looking into two GPT4 models (fairness dicussed in Appendix \ref{app: gpt-4-compare-fairness}).
GPT-4-0613 + naive RAG outperforms GPT-4-1106-preview by an average of 6.13\% pass/success rate increase on Torchdata-Github, and achieves comparable performance on more challenging Torchdata-Manual.

\paragraph{ExploraCoder vs naive RAG on API-untrained models.} 

From Table \ref{tab:Torchdata-Github-Main}, we can observe that ExploraCoder brings substantial improvements over naive RAG for both API-untrained models (GPT-3.5-turbo-0125 and GPT-4-0613), with an average absolute gains in pass@20 of 3.5\% on Torchdata-Github and 19.8\% on Torchdata-Manual.
These improvements could be attributed to ExploraCoder's potential in addressing two limitations of the naive RAG framework when handling complex API invocation subtasks: 

(1) \emph{Retrieval for complex requirement}: In the naive RAG approach, we empirically find the retriever’s ability to recall relevant APIs for comprehensive requirements becomes a bottleneck. 
ExploraCoder addresses this by adopting a divide-and-conquer strategy, identifying APIs for each explicit subtask. Additionally, ExploraCoder alleviates the need for manual hyperparameter tuning by fixing retrieval counts per subtask and dynamically adjusting subtask numbers.

\begin{table*}
    \centering
    
    \resizebox{\textwidth}{!}{

    \begin{tabular}{lcccccccc}
    \hline
    
      \multirow{2}{*}{Method}  & \multicolumn{2}{c}{\bfseries{$k=1$}} & \multicolumn{2}{c}{\bfseries{$k=5$}} & \multicolumn{2}{c}{\bfseries{$k=10$}} & \multicolumn{2}{c}{\bfseries{$k=20$}} \\
    \cmidrule(lr){2-3} \cmidrule(lr){4-5} \cmidrule(lr){6-7} \cmidrule(lr){8-9}
     & Pass& Success& Pass& Success& Pass& Success& Pass& Success\\
    \hline

    Direct Generation& 0\%& 0\%& 0\%& 0\%& 0\%& 0\%& 0\%& 0\%
\\
     DocPrompting \citeyearpar{zhou_docprompting_2023}& 0.19\%& 0.61\%& 0.89\%& 2.92\%& 1.66\%& 5.47\%& 2.81\%& 9.53\%
\\
     CAPIR \citeyearpar{ma_compositional_2024}& 3.01\%& 4.79\%& 6.75\%& 10.16\%& 8.21\%& 15.09\%& 9.66\%& 21.25\%
\\
     EpiGen \citeyearpar{li2024epigen}& 2.16\%& 5.43\%& 4.40\%& 12.33\%& 5.23\%& 15.20\%& 5.86\%& 18.46\%
\\
     \textbf{ExploraCoder (Ours)}& \textbf{7.00\%}&\textbf{ 14.8\%}& \textbf{11.54\%}& \textbf{22.88\%}& \textbf{13.84\%}& \textbf{25.4\%}& \textbf{15.67\%}& \textbf{27.56\%}
\\
     \hline
     ReAct \citeyearpar{yao2022react}& 2.00\%& 6.38\%& 2.48\% & 10.66\% & 2.95\%& 12.45\%& 3.90\%& 13.90\%
\\
     KnowAgent \citeyearpar{zhu2024knowagent}& 6.81\%& 20.54\%& 9.82\% & 22.70\% & 11.01\%& 23.29\%& 11.76\%& 23.53\%
\\
     CAPIR + Self-Repair \citeyearpar{olausson_is_2024}& 7.47\%& 15.35\%& 8.32\%& 19.08\%& 8.64\%& 20.46\%& 8.89\%& 21.66\%
\\
     \textbf{ExploraCoder* (Ours)}& \textbf{11.5\%}& \textbf{21.35\%}& \textbf{18.32\%}& \textbf{32.76\%}& \textbf{20.87\%}& \textbf{36.81\%}& \textbf{23.51\%}& \textbf{40.16\%} \\
    \hline

    \end{tabular}
    }
    \caption{Comparing ExploraCoder with advanced retrieval-based approaches using GPT3.5 on Torchdata-Manual.}
    \label{tab:Torchdata-Manual-baselines}
    \vspace{-10pt}
\end{table*}
(2) \emph{Generating code with multiple unseen APIs}: 
The complexity of coding with multiple unseen APIs lies in understanding the limited documentation and reasoning over multi-API interactions' behavior (We provide case study in Appendix \ref{app: casestudy}).
ExploraCoder mitigates this challenge by adopting a human-like exploratory programming paradigm, incrementally generating simple, reusable API invocations during CoAE, and learning extra usage knowledge from intermediate output.

\paragraph{ExploraCoder on API-pretrained model.}
We observe the API-pretrained models underperform on Torchdata-Manual, with the most competitive GPT-4-1106-preview achieving only 0.16\% in pass@1. Therefore, we use GPT-4-1106-preview on Torchdata-Manual benchmark as a proxy to further examine the effectiveness of ExploraCoder on API-pretrained models. 
Results in Table \ref{tab:Torchdata-Github-Main} shows ExploraCoder brings a substantial improvement for GPT-4-1106-preview, with an absolute pass@1 increase of 14.46\%, and it also outperforms GPT-4-1106-preview + naive RAG by 11.43\%. These results indicate that ExploraCoder is universally effective, improving models with varying levels of pretraining on relevant API knowledge.

\subsection{Experience exploitation for ExploraCoder}
\begin{figure}[ht]
    \centering
    \includegraphics[width=\columnwidth]{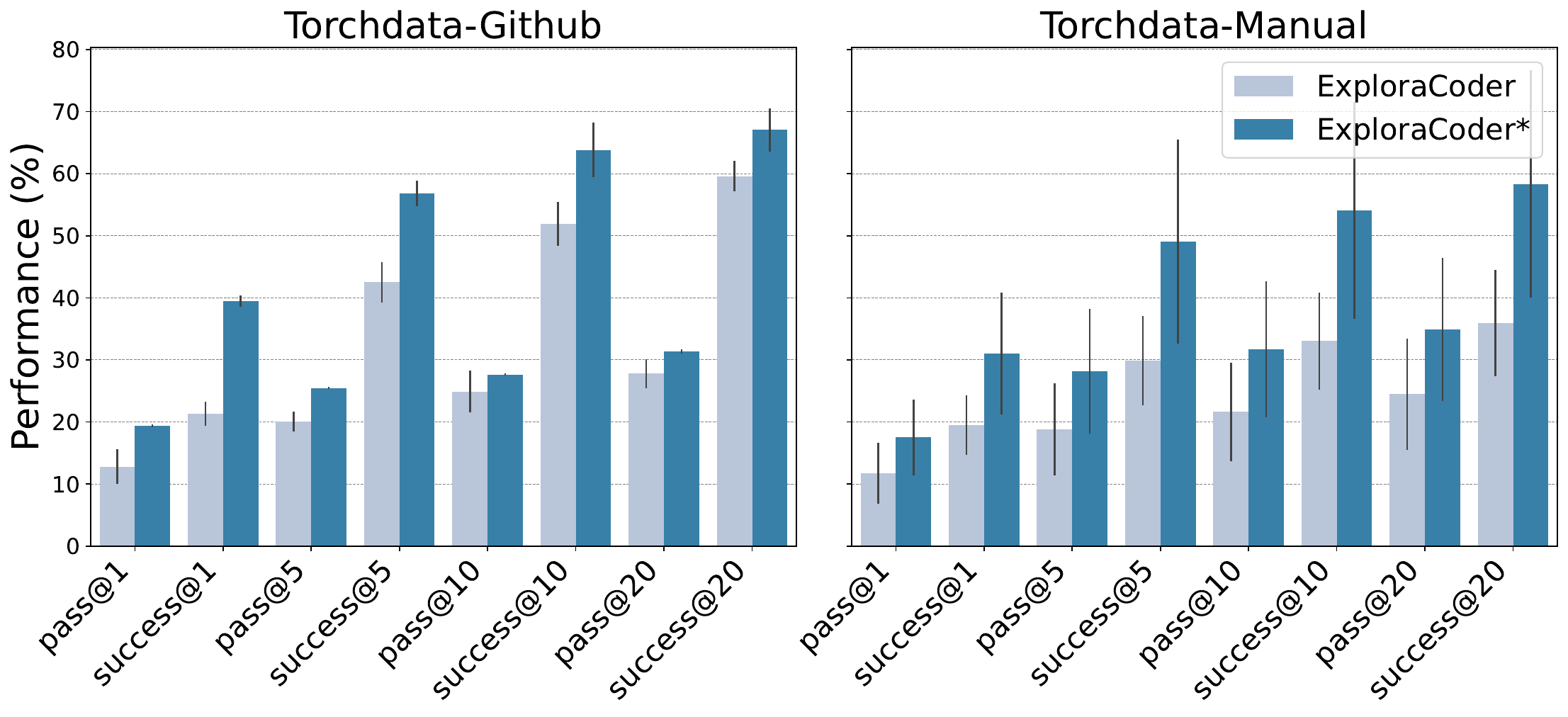}
    \captionsetup{skip=6pt}
    \caption{
    Performance comparison on the Torchdata-GitHub and Torchdata-Manual datasets across two methods (ExploraCoder and ExploraCoder*). Each bar represents the mean performance GPT-3.5-turbo-0125 and GPT-4-0613 for pass/success rate, with the range lines indicating the variation between the two models.
    }
    \label{fig:ExploraCoder_debug}
    \vspace{-10pt}
\end{figure}

In multi-API tasks, subsequent API invocations often depend on the outputs of earlier APIs. A failure in a dependent API could cascade into subsequent API invocations, regardless of whether their usage is correct. In ExploraCoder, unresolved subtasks could hinders the accuracy of the complete solution. An advancement of CoAE's step-wise code generation is the API failures can be observed in early stage. This provides the opportunity to debug on the intermediate codes, which proves to outperform debugging on full code in Section \ref{subsec:RQ3_baselines}.

To this end, we designed an enhanced ExploraCoder* by integrating a self-debug mechanism into CoAE.
When all the candidate codes are non-executable, we exploit the failed API usage experience by debugging. Figure \ref{fig:ExploraCoder_debug} shows ExploraCoder* significantly boosts the final solution's quality on two models across two benchmarks, achieving an average relative increase of 55.8\% in pass@1 and 71.3\% in success@1. More quantitative analysis of CoAE is provided in Appendix \ref{app: quantitative-analysis}.

\subsection{Comparing with related approaches}
\label{subsec:RQ3_baselines}
In this section, we further compare ExploraCoder with more advanced retrieval-based approaches. We also include Docprompting, the previously reported naive RAG framework, along with direct generation as baselines. We compare the features and computational costs in each baseline in Appendix \ref{app:compare_baselines}.

For SOTA multi-API-oriented methods, CAPIR and EpiGen, we set a fixed number of API recommendation in accordance with our $A_G$, and we directly use the subtasks generated by ExploraCoder's planning module as the preactive plannings for EpiGen. Tables \ref{tab:Torchdata-Manual-baselines} and Table \ref{tab:Torchdata-Github-baselines} show that ExploraCoder surpasses these methods by enriching the API knowledge from trial executions, compensating for the potential ambiguity in retrieved docs, achieving an absolute increase of 10.87\% in pass@10 across the two benchmarks.

\begin{table*}
    \centering
    
    \resizebox{\textwidth}{!}{

    \begin{tabular}{lcccccccc}
    \hline
    
     \multirow{2}{*}{Method}  & \multicolumn{2}{c}{\bfseries{$k=1$}} & \multicolumn{2}{c}{\bfseries{$k=5$}} & \multicolumn{2}{c}{\bfseries{$k=10$}} & \multicolumn{2}{c}{\bfseries{$k=20$}} \\
    \cmidrule(lr){2-3} \cmidrule(lr){4-5} \cmidrule(lr){6-7} \cmidrule(lr){8-9}
     & Pass& Success& Pass& Success& Pass& Success& Pass& Success\\
    \hline

    Direct Generation& 1.70\%& 2.09\%& 5.54\%& 6.95\%& 7.28\%& 9.64\%& 8.00\%& 11.90\%\\
     DocPrompting \citeyearpar{zhou_docprompting_2023}& 6.00\%& 10.57\%& 10.55\%& 24.00\%& 14.67\%& 32.50\%& 20.83\%& 40.81\%\\
     
     CAPIR \citeyearpar{ma_compositional_2024}& 5.90\%& 10.47\%& 14.59\%& 27.08\%& 18.60\%& 37.19\%& 23.52\%& 47.43\%\\
     
     EpiGen \citeyearpar{li2024epigen}& 8.57\%& 18.95\%& 14.63\%& 35.61\%& 17.24\%& 41.67\%& 19.61\%& 47.62\%\\
     \textbf{ExploraCoder (Ours)}& \textbf{10.19\%}& \textbf{19.50\%}& \textbf{18.64\%}& \textbf{39.39\%}& \textbf{21.67\%}& \textbf{48.56\%}& \textbf{25.62\%}& \textbf{57.30\%} \\
     \hline
     
     ReAct \citeyearpar{yao2022react}& 10.19\%& 27.90\%& 10.95\% & 33.06\% & 11.90\%& 33.88\%& 13.81\%& 34.00\%
\\
     KnowAgent \citeyearpar{zhu2024knowagent}& 14.67\%& 25.81\%& 15.99\% & 30.68\% & 16.00\%& 31.90\%& 16.00\%& 33.81\%
\\
     CAPIR + Self-Repair \citeyearpar{olausson_is_2024}& 16.47\%& 22.10\%& 21.04\%& 29.70\%& 21.75\%& 32.20\%& 22.00\%& 33.90\%\\
     \textbf{ExploraCoder* (Ours)}& \textbf{19.24\%}& \textbf{38.66\%}& \textbf{25.41\%}& \textbf{54.93\%}& \textbf{27.64\%}& \textbf{59.56\%}& \textbf{31.62\%}& \textbf{63.71\%}\\
    \hline

    \end{tabular}
    }
    \caption{Comparing ExploraCoder with advanced retrieval-based approaches using GPT3.5 on Torchdata-Github.}
    \label{tab:Torchdata-Github-baselines}
\end{table*}
To compare ExploraCoder* with other debug-enhanced methods, we first adapted a SOTA debugging framework, Self-Repair\footnote{To ensure the fairness in debug iteration budget, for each problem, if ExploraCoder generates $n$ plans, enabling debugging in up to $n$ CoAE steps, we set the iteration budget for Self-Repair in that problem to $n$ accordingly.}, by augmenting the LLM with API knowledge retrieved by CAPIR throughout its iterative 2-stage generation.
We also compare with two agentic framework, ReAct and KnowAgent, specifically designed for reactive knowledge retrieval during the solution generation. We sample the candidates from their `Finish` action, which derives final solutions based on the interleaving retrieval and debugging trajectory.

Table \ref{tab:Torchdata-Manual-baselines} and Table \ref{tab:Torchdata-Github-baselines} shows that while these iterative/agentic methods benefit from the debugging, the overall improvement, especially in bigger k, remains limited. This could be due to the limitation of reactive planning, which is bug-driven (Appendix \ref{app:compare_baselines})
and lacks systematic understandings of API knowledge for diverse solution implementations. 
ExploraCoder*, through
intermediary debugging on simpler subtasks, exhibits a significant pass@10 increase over 9.06\%.
Even ExploraCoder achieves a comparable performance on Torchdata-Github and surpasses them on the more complex Torchdata-Manual. This highlights our superior design in uniquely enforcing a step-wise code construction workflow and iteratively enhancing the preactive plans with exploratory knowledge.

\subsection{Ablation study}
\begin{table}
    \centering
    
    \resizebox{\columnwidth}{!}{

    \begin{tabular}{lcccc}
    \hline
    
     \multirow{2}{*}{Method}  & \multicolumn{2}{c}{\bfseries{$k=1$}} & \multicolumn{2}{c}{\bfseries{$k=10$}}\\
    \cmidrule(lr){2-3} \cmidrule(lr){4-5}
     & Pass& Success& Pass& Success\\
    \hline
    \rowcolor{gray!15}
    ExploraCoder*& 11.5\%& 21.35\%
& 20.87\%& 36.81\%
\\
     w/o Self-Debug& 7.00\%& 14.8\%
& 13.84\%& 25.4\%
\\
    \rowcolor{gray!15}
     w/o Lib-ICL& 7.64\%& 15.73\%
& 14.17\%& 29.77\%
\\
     w/o CoAE& 1.22\%& 2.21\%
& 7.34\%& 13.38\%
\\
    \rowcolor{gray!15}
     w/o Selection& 4.12\%& 13.33\%& 9.46\%& 26.38\%\\
    \hline

    \end{tabular}
    }
    \caption{Ablation study for ExploraCoder framework using GPT3.5 on Torchdata-Manual.}
    \label{tab:Torchdata-Manual-Ablation}
    \vspace{-15pt}
\end{table}
We further conducted an ablation study on our best-performing framework ExploraCoder* in Table \ref{tab:Torchdata-Manual-Ablation}. We experiment on the challenging Torchdata-Manual benchmark using GPT-3.5-turbo-0125.

As discussed earlier, self-debugging intermediate execution failure effectively improves ExploraCoder's performance. This also suggests ExploraCoder may further benefit from dynamically generated testbed for intermediate code, which we will leave as exciting future work to explore.

We ablate the in-context learning of library-level knowledge (w/o Lib-ICL), removing the few-shot planner $\mathcal{D}$ and library introduction $s$, and let the model plan API invocation subtasks based soley on its commonsense knowledge. The performance decline observed could be attributed to the misalignment between planned subtasks and API granularity. Since overly coarse-grained subtasks introduce complexity, while incorrect subtasks that cannot be solved by any APIs increases the hallucination rates \citep{liu2024apihallu,tian2024codehalu}.

We ablate the CoAE (w/o CoAE) by providing all the retrieved API documentation throughout ExploraCoder's process to the generator, and prompt it to end-to-end generate final solution. We find that the performance significantly drop to 1.22\% in pass@1. This suggests (1) modern generators still lack adequate in-context reasoning ability to handle multiple unseen API invocations, and (2) API documentation could be insufficient, leading to hallucinated invocations. This highlights the need for intermediate execution or debug to gain more usage insights.

We further ablate a critical step within CoAE by removing the experience selection process (w/o selection). In this variant, candidate selection is randomized, disregarding executability signals. We find the success rate remains reasonably well, and the pass rate declines. A possible explanation is ExploraCoder degenerates into exploring low-quality API invocation chains with limited usage insights for fully accurate final solution.

\section{Conclusion}
We present ExploraCoder, a novel code generation framework for LLMs to generate multiple unseen API invocations through planning API-related subtasks and experimenting with each subtask in a novel chain-of-API-exploration. Experiments on our newly constructed benchmark and an existing benchmark demonstrates ExplroaCoder's superior performance over competitive approaches.

\section{Limitations}
ExploraCoder's effectiveness relies on the underlying LLM's capabilities in handling long contexts and capturing API usage knowledge. Although our experiments show strong performance with both GPT-3.5 and GPT-4, small models with weak capability could exhibit less effectiveness on our complex multi-API-generation tasks. This dependency means that the framework's performance is inherently bounded by the LLM's capabilities. However, the rapid advancement in LLM development suggests this limitation may become less significant over time.

The framework assumes the availability of NL documentation, which may limit its effectiveness when dealing with overly incomplete, ambiguous, or erroneous API documentation. In our experiments, we simulated real-world scenarios by masking detailed parameter explanations and usage examples from the well-maintained torchdata documentation, approximating the minimal documentation typically available for newly introduced or privately maintained libraries. While this setting demonstrates ExploraCoder's robustness with minimal API descriptions, future work could explore integrating additional knowledge sources, such as API client code or community discussions, to supplement insufficient documentation.

A promising improvement direction shared by ExploraCoder and related approaches is an early termination mechanism in the iterative generation workflow. When encountering particularly challenging problems where API exploration consistently fails, the system continues attempting solutions, potentially consuming unnecessary computational resources. The development of intelligent stopping criteria that can identify unsolvable problems or determine when further exploration would be unproductive represents an important direction for future research.

\bibliography{exploracoder}

\appendix

\section{Appendix}

\subsection{Comparing ExploraCoder with related approaches}
\label{app:compare_baselines}
We present a feature comparison in Table \ref{tab:baseline_features} and a detailed breakdown of the computational overhead (model call and token consumption) in Table \ref{tab:baseline_computational_cost}.
\subsubsection{Feature Comparison}
ExploraCoder introduces two key innovations in LLM-based code generation: exploratory planning and step-wise code construction. (1) Firstly, traditional approaches generally follow either "preactive" planning (based on prior knowledge) or "reactive" planning (based on environmental feedback). Preactive planning, such as CoT prompting, can suffer from hallucinations when handling complex or out-of-distribution tasks. Reactive planning, common in agent-based systems, often lacks systematic consideration and controllability \citep{xia2024agentlessdemystifyingllmbasedsoftware}. ExploraCoder bridges this gap by introducing exploratory planning, which enhances preactive plans with step-wise environmental feedback to mitigate hallucinations while maintaining systematic control.
(2) Secondly, most existing work conduct end-to-end code generations/modification. We propose a step-wise code construction workflow to generate partial code for a simple subtask based on the planning instructions. \citet{le_codechain_2024} exhibits a similar idea of preactively planning a series of reusable functions for simple self-contained code generation, while it fails to leverage the step-wise execution signal from partial codes and it is also not applicable to more complex programming scenario like multi-unseen-API invocations.

\paragraph{ExploraCoder vs. Existing library-oriented approaches.}
While existing library-oriented approaches (DocPrompting, CAPIR, EpiGen) primarily focus on API-docs retrieval quality, ExploraCoder addresses the fundamental limitations in LLMs' multi-API reasoning capabilities and documentation ambiguity. Through a novel Chain of API Exploration, ExploraCoder iteratively collects execution information to resolve API usage ambiguities. For instance, in task\_175, traditional preactive planning in EpiGen hallucinated about the parameter types in "LineReader". ExploraCoder resolves such issues by executing multiple candidate implementations for the "read the lines" subtask and filtering out incorrect API usage patterns, thereby acquiring additional API knowledge that cannot be derived from documentation alone.

\definecolor{myteal}{RGB}{0,128,128}      
\definecolor{myblue}{RGB}{65,105,225}     
\definecolor{mygreen}{RGB}{34,139,34}     

\begin{table*}[t]
\centering
\small
\begin{tabular}{lclcccc}
\hline
Method& API Retrival& Planning& Step-wise Code Construction& Debugging & Manual Requirement\\
\hline
DocPrompting& \yes & \no & \no&\no&-\\

CAPIR & \yes & \textbf{\textcolor{myblue}{Preactive}$^\dagger$} & \no &  \no &-\\

EpiGen& \yes& \textbf{\textcolor{myblue}{Preactive}} & \no&\no&-\\
ExploraCoder&\yes & \textbf{\textcolor{myteal}{Exploratory}} & \yes &\no&-\\
\midrule
Self-Repair& \no& \textbf{\textcolor{mygreen}{Reactive}}& \no &\yes &-\\
ReAct&\yes & \textbf{\textcolor{mygreen}{Reactive}} & \no &\yes & \textbf{\textcolor{gray}{Agentic traj. fewshot}}\\
KnowAgent&\yes & \textbf{\textcolor{mygreen}{Reactive}} &\no& \yes& \textbf{\textcolor{gray}{Agentic traj. fewshot}}\\
CAPIR + Self-Repair& \yes& \textbf{\textcolor{mygreen}{Reactive}$^\dagger$}& \no&\yes&-\\
ExploraCoder*& \yes& \textbf{\textcolor{myteal}{Exploratory}}& \yes&\yes&-\\
\hline        
\end{tabular}
\caption{Features in each retrieval-based baselines. $\dagger$: CAPIR focuses exclusively on planning for the retrieval phase, without addressing the generation phase.
}
\label{tab:baseline_features}
\end{table*}
\begin{table*}
\centering
\resizebox{\textwidth}{!}{
\begin{tabular}{@{}lccccc@{}}
\toprule
Model                & Pre-Processing Calls & Code Generation Calls& Overall Model Calls & Tokens & Pass@10 \\ \midrule
DocPrompting            & 0                      & 1                       & 1                   & 10k                   & 8.15\%\\
CAPIR               & n+2                & 1                       & 3+n             & 18k                   & 11.23\%\\
EpiGen              & n+2                & 1                       & 3+n             & 18k                    & 13.40\%\\
ExploraCoder        & n+2                & n+1                & 3+2n           & 56k                    & \underline{17.76\%} \\
CAPIR + Self-Repair & n+2                & 1.5n+2& 4+2.5n& 70k& 15.19\%\\
ExploraCoder*       & n+2                & 2.6n+1& 3+3.6n& 95k& \textbf{24.26\%} \\ 
ReAct & - & -& 2N+2& \underline{112k} & 7.43\% \\
KnowAgent & - & - & 3N+3& \textbf{143k} & 13.51\% \\
\bottomrule
\end{tabular}
}
\caption{Computational costs and performance in each retrieval-based baselines.}
\label{tab:baseline_computational_cost}
\end{table*}

\paragraph{ExploraCoder vs. iterative-debugging.}
ExploraCoder's step-wise code construction offers significant advantages over existing iterative-debugging approaches. While methods like CAPIR+Self-Repair employ bug-driven reactive planning on complete code solutions, ExploraCoder* debugs simpler subtasks at earlier stages, preventing the accumulation of complex errors. We observe in case study (Appendix \ref{app: casestudy}) that CAPIR+Self-Repair repeatedly attempts to fix a buggy code that deviates substantially from the correct solution, continuing until it exhausts its iteration budget.

\paragraph{ExploraCoder vs. Agent-style frameworks.}
ExploraCoder differs from agent-style frameworks in three crucial aspects:

1) \textbf{Structured Workflow:} Unlike agent frameworks with undeterministic actions, ExploraCoder implements a pinned step-wise code construction workflow (exploratory programming) that experimentally performs well with multi-API invocations. Most code agents, eg. Swe-Agent \citep{yang2024sweagentagentcomputerinterfacesenable}, MetaGPT \citep{hong2024metagptmetaprogrammingmultiagent}, conduct end-to-end code construction/modification, just like what we discussed with Self-Repair. In our agentic baseline implementations (Appendix \ref{app:agent implementation}), we borrow the idea of ExploraCoder and prompt React and KnowAgent to generate partial code at each step. Our empirical results show that enforcement of step-wise code generation in agentic workflow is often unstable and uncontrollable, which aligns with the suggestions with Agentless \citep{xia2024agentlessdemystifyingllmbasedsoftware}.

2) \textbf{Systematic Planning:} ExploraCoder's exploratory planning maintains a comprehensive view of task dependencies, preventing common pitfalls seen in reactive planning. For example, in task\_124 (CSV loading from compressed files), reactive agents often overlook crucial steps like decompression, leading to inefficient API retrieval cycles. ExploraCoder's exploratory planning first systematically breaks down such tasks into logical, dependent steps. Then it also leverages the environmental feedback to enhance the next-step plannings with additional API knowledge.

3) \textbf{Efficiency and Accessibility:} Agent-based approaches require high-quality few-shot examples for reasoning trajectories, which are often impractical when working with new libraries. We empirically find ReAct and KnowAgent performance deteriorize when we remove the examplary trajectory or provide an OOD trajectory on other libraries (See implementation details in Appendix \ref{app:agent implementation}). Additionally, they tend to be token-inefficient due to potential deterioration into recursive or meaningless actions when facing noisy observations.

\subsubsection{Computational Comparison}
From Table \ref{tab:baseline_computational_cost} we can observe that ExploraCoder is cost-efficient. When Compared to CAPIR+Self-Repair and two agentic methods, ExploraCoder achieves higher performance with fewer model calls and lower token consumption. Additionally, ExploraCoder* achieves a 59.7\% performance improvement over CAPIR+Self-Repair with a 35.71\% increase in token consumption and a 44\% increase in model calls, demonstrating that the performance gains significantly outweigh the proportional increase in resource consumption.

\textbf{Calculation details:} For model calls, we provide clear analytical expressions based on n, the number of decomposed subtasks. For token consumption, we randomly sampled a task subset with n=8 (the mean value of n in our datasets), and generated 20 candidate final solutions, then calculated the average token consumption per task.
Note that the agent-based methods' action is uncontrollable, and their model calls cannot be mapped to a preactively determined n, therefore we use a different N to represent its iteration steps. We empirically observe N>n in most tasks.
For the two non-agent approaches involving self-debug mechanisms, Self-Repair and ExploraCoder*, debugging rounds is not deterministic. Therefore we use the formulated expectation based on the empirically observed debug rate in our experiments. Having the probability of the two methods conducting debug as 
\(p_1\) (ExploraCoder*) and \(p_2\) (Self-repair), their expectation of model call can be formulated as \((1+5p_1)n+1\) and \(2p_2n+2\). Notably, \(p_2=0.75\) is significantly higher than \(p_1=0.32\) across two benchmark. This debug rate difference arises because ExploraCoder* focuses on debugging simple intermediate subtasks (which are generally less error-prone), while Self-Repair always attempts to debug a complete solutions (which often fail to repair successfully, triggering additional debugging iterations up to the budget limit).

\subsubsection{Implementation details of KnowAgent and ReAct}
We use the official code and prompt released to implement ReAct \citep{yao2022react} and KnowAgent \citep{zhu2024knowagent}:
\label{app:agent implementation}
\paragraph{Action Space:} In our experiment, we abstract the capability of ExploraCoder* and design the following actions/tools for ReAct and KnowAgent:
\begin{enumerate}
\item \texttt{Retrieve[target\_functionality]}: Query Torchdata API documentation for a specific functionality, returning top-k relevant APIs.

\item \texttt{Write\_and\_Execute[code]}: Generate/Debug then execute an in-progress code snippet to test partial functionality. The execution information will be returned.

\item \texttt{Finish[code]}: Write the complete code solution that solves the coding task based on reasoning trajectroy.
\end{enumerate}

\paragraph{Trajectory example:}
We manually crafted one long trajectory example of the agent-style Torchdata task-solving process across 8 reasoning steps, showing the interleaving trace of API retrieval and code generation, code debug, and more API retrieval, etc.
While this enables the agent to understand the expected reasoning flow, we note this manual involvement is expensive in real-world deployment, especially for newly-introduced libraries. We empirically observe their performance deteriorize when we remove the examplary trajectory or provide an OOD trajectory on other libraries. 

\paragraph{Reasoning steps \& model call budget:} To ensure fair comparison given our tasks' complexity, we extended the reasoning step budget of React and KnowAgent to 16 on Torchdata-Github and 21 on Torchdata-Manual (vs. original 10), enabling them to initiate 32/42 and 48/63 model calls to perform analysis, planning, and write/debug code snippets. Table \ref{tab:baseline_computational_cost} shows that the token consumption are significantly higher than non-agent baselines.

\begin{figure*}[th]
    \centering
    \begin{subfigure}[b]{0.245\textwidth} 
        \centering
        \includegraphics[width=\textwidth]{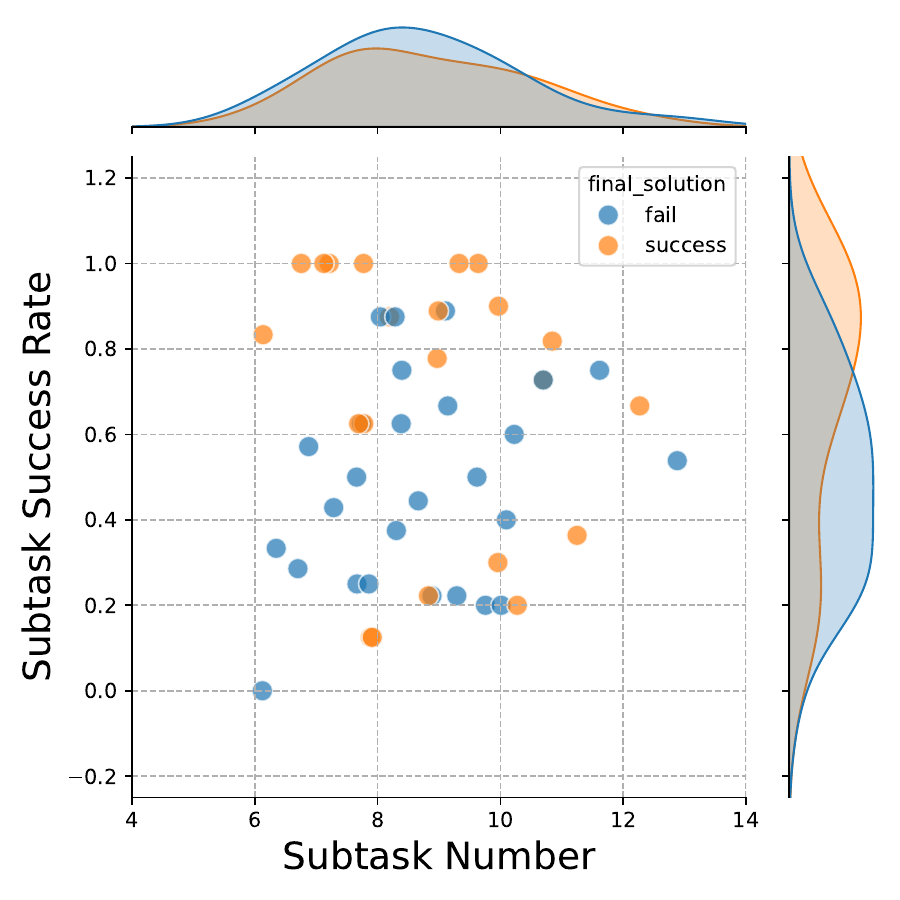}
        \caption{success-fail division\\ for final solution in\\ ExploraCoder}
        \label{fig: a}
    \end{subfigure}
    \hfill 
    \begin{subfigure}[b]{0.245\textwidth} 
        \centering
        \includegraphics[width=\textwidth]{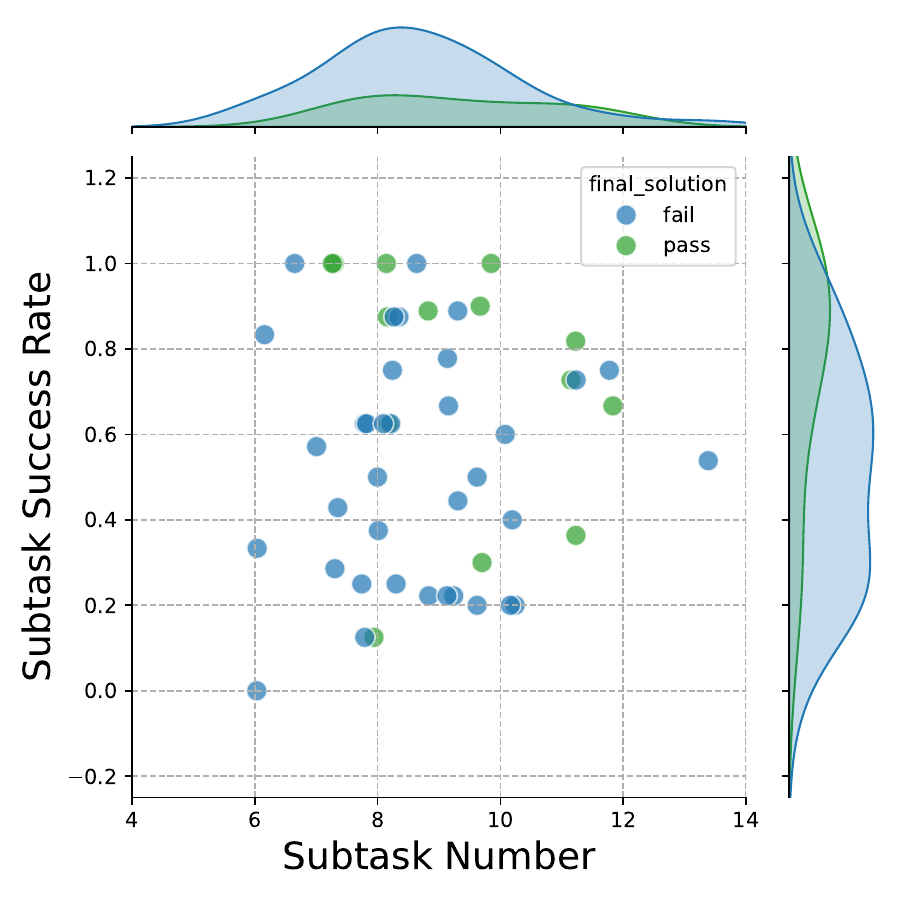} 
        \caption{pass-fail division\\ for final solution in\\ ExploraCoder}
        \label{fig: b}
    \end{subfigure}
    \hfill 
    \begin{subfigure}[b]{0.245\textwidth} 
        \centering
        \includegraphics[width=\textwidth]{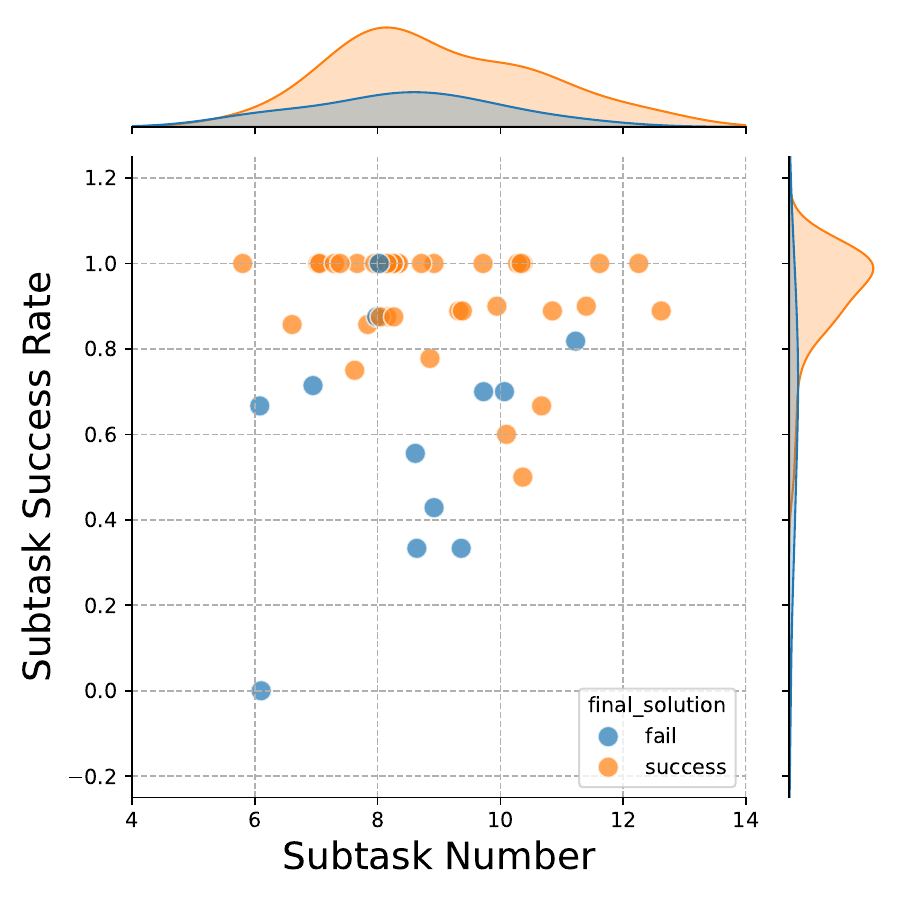} %
        \caption{success-fail division\\ for final solution in\\ ExploraCoder*}
        \label{fig: c}
    \end{subfigure}
    \hfill 
    \begin{subfigure}[b]{0.245\textwidth} 
        \centering
        \includegraphics[width=\textwidth]{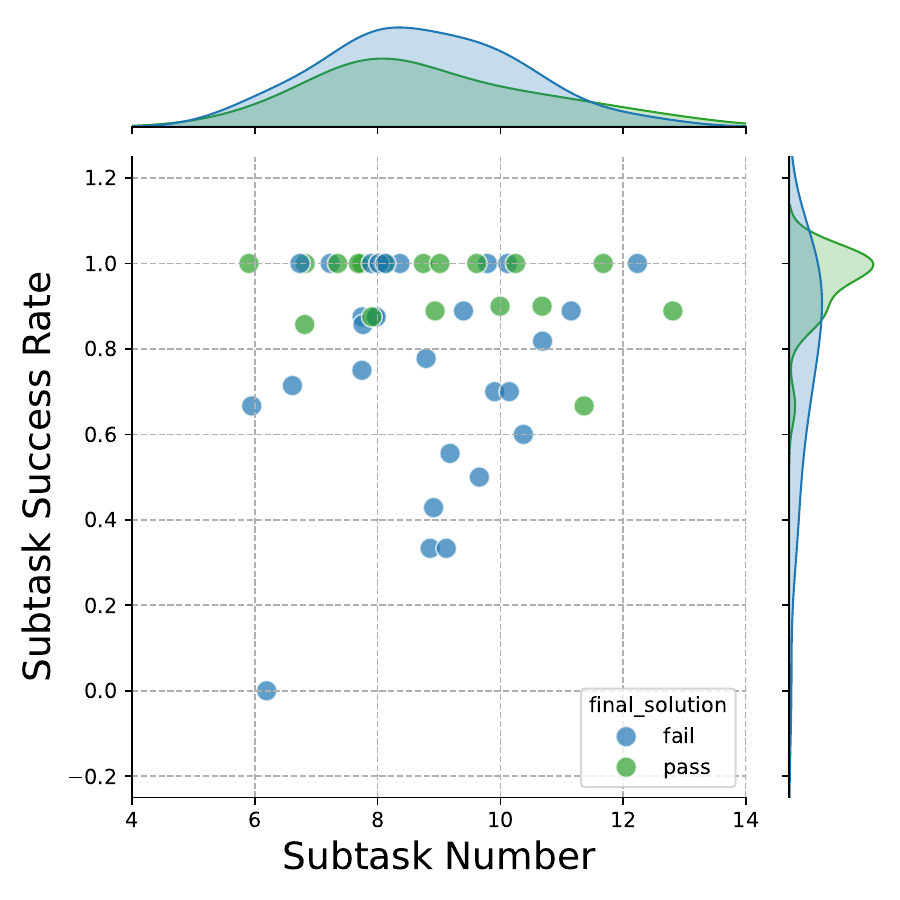} 
        \caption{pass-fail division\\ for final solution in\\ ExploraCoder*}
        \label{fig: d}
    \end{subfigure}
    \caption{Correlation between the quality of CoAE subtasks and final solutions}
    \label{fig: ExploraCoder_scatterplot}
\end{figure*}
\subsection{Comparing ExploraCoder with repeated-sampling}

Inferior baselines sometimes have simpler framework designs thus costing fewer tokens. Their performance could naturally be improved through scaling up the inference-time computation in the sampling phase. We investigate whether repeated-sampling token-efficient methods like CAPIR can achieve competitive performance comparable to token-intensive approaches like ExploraCoder* with equivalent token budgets. Experiments on 50 randomly sampled problems from Torchdata-Manual in Table \ref{tab:repeated_sampling} shows that:
\begin{table}[ht]
\centering
\small
\resizebox{\columnwidth}{!}{
\begin{tabular}{lcc|ccccc}
\hline
\multirow{2}{*}{Method} & \multicolumn{2}{c|}{Basic Metrics} & \multicolumn{5}{c}{Extended Metrics} \\
\cline{2-8}
 & Pass@10& Pass@20& Pass@50 & Pass@90 & Pass@100& Pass@120 & Pass@130 \\
\hline
CAPIR & 8.48\%& 9.70\%& 11.84\% & 13.33\% & 13.52\%& 13.71\% & 13.72\% \\
\arrayrulecolor{gray!15}\hhline{~~~|>{\arrayrulecolor{black}}-----}
ExploraCoder & 16.07\%& 17.55\%& \multicolumn{5}{>{\columncolor{gray!15}}c}{\textit{not evaluated}} \\
\arrayrulecolor{gray!15}\hhline{~~~|>{\arrayrulecolor{black}}-----}
ExploraCoder* & 24.99\%& 27.22\%& \multicolumn{5}{>{\columncolor{gray!15}}c}{\textit{not evaluated}} \\
\hline
\end{tabular}
}
\caption{Pass@k performance comparison.}
\label{tab:repeated_sampling}
\end{table}

\paragraph{1) Efficient method like CAPIR doesn’t neccessarily yield competitive results by repeated-sampling.} Under same level of token consumption \footnote{Token calculation details:
As shown in Table \ref{tab:repeated_sampling}, the fixed token difference between CAPIR and ExploraCoder is around 38k($\approx$ \$0.038), let a marginal token consumption per sample be 450 (empirically on Torchdata-Manual), we can equalize the token consumption between CAPIR and ExploraCoder when CAPIR have 84 more samples than ExploraCoder, which means token(CAPIR, k=100) $\approx$ token(ExploraCoder, k=20)

}, CAPIR k=100 (pass=13.52\%) underperform than ExploraCoder k=20 (pass=17.55\%). And it even underperfrorms than ExploraCoder with k=10 (pass=16.07\%).

\paragraph{2) The improvement in pass rate plateaus as the sampling number k increases.} CAPIR’s pass rate barely improve when we scale the k from 100 to 130. This aligns with \citet{brown2024largelanguagemonkeysscaling}'s findings that inference-time sampling typically follows logarithmic scaling laws. This indicate inferior methods like CAPIR should take even much larger computational costs to possibly achieve adequate performance with ExploraCoder. 

\paragraph{3) Large repeated-sampling number present fundamental fairness and practicality concerns.}
While evaluating methods under equivalent computational budgets provides valuable insight, we’d like to suggest that raising large sampling numbers (eg. k$\ge$100) creates practical challenges for candidate verification, which limits its real-world applicability. \citet{brown2024largelanguagemonkeysscaling} indicates that despite the effectiveness of repeated sampling for correct answer, it is often hard for users (verifiers) to verify a candidate solution from the large volumes of samples. For example, test suites are often inaccessible during development of new functionalities \citep{testcasehard1}. Furthermore, managing a large number of candidate solutions does not align with typical development workflows.

Our experiments demonstrate that ExploraCoder standsout in two critical aspects: (1) its \textbf{ability to generate correct solutions with a manageable sampling budget}, and (2) its \textbf{superior token efficiency} compared to computation-scaling variants of CAPIR (including both w/ repeated-sampling and w/ Self-Repair workflow). These results highlight ExploraCoder's practical advantages in real-world applications, where both solution quality and computational efficiency are essential considerations.

\subsection{Quantitative analysis}

\subsubsection{The effectiveness of CoAE in ExploraCoder}
\label{app: quantitative-analysis}
ExploraCoder leverages API invocation experience from CoAE to enhance the quality of final solution generation. Intuitively, the quality of exploration subtasks within CoAE is closely related to the quality of the final solutions.

To explore the pass/success rate relationship between CoAE subtasks and final solutions in ExploraCoder, we conducted a quantitative analysis, examining how the number of CoAE subtasks and their success rates affect the pass rate and overall success rate of the final solutions. We illustrate their correlation using results from the best-performing base model, GPT-4-0613, on our Torchdata-Manual benchmark.

From Figure \ref{fig: a} to \ref{fig: d}, we observe that both the pass rate and success rate of the final solutions positively correlate with the CoAE subtask success rate. Subtasks with higher success rates, particularly those with a success rate of 1, are more likely to generate successful or passing final solutions. 
Interestingly, the number of subtasks doesn’t appear to have a significant direct impact. However, as shown in Figures \ref{fig: a} and \ref{fig: b}, without intermediate self-debug, problems with a higher subtask number (ranging from 10 to 13) tend to have lower subtask success rate as the subtask number increase. This may be due to the increased complexity of inter-task API interactions. When the self-debug mechanism is introduced in ExploraCoder*, we observe in Figures \ref{fig: c} and \ref{fig: d} a notable improvement in the overall subtask success rate, even for cases with higher subtask numbers. This leads to more successful and passing final solutions. The improvement can be attributed to ExploraCoder’s ability to correct typos and simple API interaction errors in each subtask, thereby gaining richer API usage experience and exploiting it to the final solution generation.

\subsubsection{The effectiveness of task planning in ExploraCoder}
Although It is hard to directly quantify the quality of decomposed tasks' granularity, we can evaluate it indirectly by calculating the number of APIs included in each subtask, since our design aims to ensure each decomposed subtask involves 1-2 API explorations, so that it's easy enough to be solved. 

As shown in Table \ref{tab:task_seg}, the average number of decomposed subtasks by GPT-3.5 is closely aligned with the average number of API invloved across two datasets. This indicates that the decomposition strategy effectively achieves the desired granularity. The ExploraCoder’s overall performance also indicates the effectiveness of our task planning.
\begin{table}[ht]
\centering
\resizebox{\columnwidth}{!}{
\begin{tabular}{lccc}
\hline
 & \#API & \#decomposed subtask & API per subtask \\
\hline
Torchdata-Github & 4.26 & 4.06 & 1.04 \\
Torchdata-Manual & 9.94 & 8.22 & 1.21 \\
\hline
\end{tabular}
}
\caption{Summary of decomposed subtask statistics.}
\label{tab:task_seg}
\end{table}

\subsection{Construction details of Torchdata-Manual}
\label{app: TorchdataManual_construction}
\begin{table}[t]
\centering
\resizebox{\columnwidth}{!}{
\renewcommand{\arraystretch}{1.2}
\begin{tabular}{lcccc}
\hline 
Benchmarks      & \begin{tabular}[c]{@{}c@{}}Num.\\samples\end{tabular} & \begin{tabular}[c]{@{}c@{}}Num.\\APIs\end{tabular} &  \begin{tabular}[c]{@{}c@{}}Num.\\Invoc.\end{tabular}& \begin{tabular}[c]{@{}c@{}}Volume of\\doc pool\end{tabular} \\
\hline 
Torchdata-Github   & 50 & 3-8  &   3-8& 228 \\
Torchdata-Manual    & 100 & 8-14&   8-21& 228 \\                                 
\hline              
\end{tabular}
}
\caption{Statistical Summary of two Torchdata-based benchmarks. Num. APIs reports the range of distinct APIs involved in each sample. Num. Invoc. reports the range of API invocations in the samples' canonical solution. Volume of the doc pool refers to the number of API documents provided by the library, which also represents the size of the search space during API retrieval.}
\label{tab:benchmarks}
\end{table}
The Torchdata-Manual benchmark is designed to provide complex programming problems that require the use of multiple Torchdata APIs. It follows the style of prior unseen library benchamarks \citep{zan_private-library-oriented_2023,ma_compositional_2024} , consisting of a natural language task description, code context, canonical solutions, and test cases. The construction process is outlined as follows:

\textbf{Torchdata API Selection.}
We first curated a subset of APIs from the complete Torchdata API pool. For each problem, we randomly sampled 15 APIs from this subset, ensuring that the selected group of APIs differed from those used in previous tasks. This process helped ensure a more balanced distribution of the Torchdata APIs and maintained the variety among problems. In total, 200 groups of 15 unique APIs were selected.

\textbf{Manual Construction of Example Programming Tasks.}
Two long-sequence API problems were manually written to serve as few-shot demonstration for the next step. Specifically, we observed and analyzed the programming problems in Torchdata-Github and manually integrated the functional requirements of several tasks while ensuring logical consistency. By combining relatively simple, real-world programming tasks to construct more complex example tasks, we believe that these examples are meaningful and representative.

\textbf{LLM based Craft Generation.}
We leverage GPT-4o, which has been trained on Torchdata knowledge, to craft some for programming problems for inspiration. Specifically, we provided the 2-shot demonstration and the documentation for the 15 APIs in each group, and tasked the GPT-4o with generating a programming problem that incorporated as many APIs as possible. This resulted in 200 initial problem drafts.

\textbf{Manual Curation of Programming Problems.}
We manually filter out reasonable problem requirements from the drafts. Based on these filtered drafts, we then rewrote high-quality, coherent problems. In total, 50 programming problems were constructed.

\textbf{Expert Review.}
Finally, we invited two Python programmers, each with four years of experience, to review the dataset and suggest adjustments. Specifically, we ask the experts to examine on 4 aspect of the crafted programming problems (1) The executability of the canonical solution, (2) The intuitiveness of the API usage, (3) The rigor of the test cases, (3) The meaningfulness of the task requirements. If any issues were identified in these aspects, the experts discussed them with the task creators and revised the tasks accordingly. This step ensured the overall quality and correctness of the benchmark. All participants were  compensated adequately, with payment aligned to ethical standards and appropriate to their demographic and region. And we ensure that there is no ethical issue involved in our data construction process.

\subsection{Generalizabiliy of 
ExploraCoder on More Unseen Libraries Settings}
\label{app:monkbeateval-baselines}
\begin{table*}
    \centering
    
    \resizebox{\textwidth}{!}{
    \begin{tabular}{lcccccccc}
    \hline
    
      \multirow{2}{*}{Method}  & \multicolumn{2}{c}{\bfseries{$k=1$}} & \multicolumn{2}{c}{\bfseries{$k=5$}} & \multicolumn{2}{c}{\bfseries{$k=10$}} & \multicolumn{2}{c}{\bfseries{$k=20$}} \\
    \cmidrule(lr){2-3} \cmidrule(lr){4-5} \cmidrule(lr){6-7} \cmidrule(lr){8-9}
     & Pass& Success& Pass& Success& Pass& Success& Pass& Success\\
    \hline

    Direct Generation& 3.40\%& 3.89\%& 6.67\%& 8.34\%& 7.80\%& 10.56\%& 8.16\%& 12.15\%\\
     DocPrompting \citeyearpar{zhou_docprompting_2023}& 8.00\%& 21.61\%& 14.01\%& 40.85\%& 18.30\%& 48.84\%& 19.70\%& 54.30\%\\
     EpiGen \citeyearpar{li2024epigen}& 7.77\%& 23.13\%& 14.41\%& 36.47\%& 17.86\%& 41.47\%& 20.30\%& 46.54\%\\
     \textbf{ExploraCoder (Ours)}& \textbf{19.73\%}&\textbf{59.47\%}& \textbf{25.75\%}& \textbf{67.36\%}& \textbf{27.36\%}& \textbf{68.76\%}& \textbf{28.47\%}& \textbf{69.39\%} \\
     \hline
     CAPIR + Self-Repair \citeyearpar{olausson_is_2024}& 13.90\%& 33.14\%& 20.78\%& 50.95\%& 23.64\%& 56.59\%& 24.49\%& 59.09\%\\
     \textbf{ExploraCoder* (Ours)}& \textbf{21.33\%}& \textbf{74.87\%}& \textbf{26.73\%}& \textbf{79.50\%}& \textbf{28.87\%}& \textbf{80.07\%}& \textbf{30.39\%}& \textbf{80.56\%} \\
    \hline

    \end{tabular}
    }
    \caption{Comparing ExploraCoder with advanced retrieval-based approaches using GPT3.5 on MonkBeatEval.}
    \label{tab:MonkBeatEval-baselines}
    \vspace{-10pt}
\end{table*} 
While Torchdata has become an established practice for evaluating unseen library code generation \citep{zan_private-library-oriented_2023,zhang2023toolcoder,ma_compositional_2024}, it is important to assess ExploraCoder's effectiveness across a broader range of libraries. However, as noted by \citet{zan_private-library-oriented_2023}, it is particularly challenging to discover a suitable library like TorchData in open-source communities. 

To enable rigorous evaluation of ExploraCoder's generalizability, we introduce MonkBeatEval, a new multi-library benchmark constructed specifically for testing performance on unseen libraries.

\subsubsection{Construction of MonkBeatEval}
\textbf{Library Creation.} We developed two pseudo-libraries (Monkey and BeatNum) by applying bidirectional transformation mappings to Pandas and NumPy APIs, using the methodology\footnote{Zan provide in their paper a set of transformations rules between Pandas/Numpy and Monkey/BeatNum} from \citet{zan_private-library-oriented_2023}, eg. "pandas.iterrows"$\leftrightarrow$"monkey.traversal”. This ensures the APIs are novel to LLMs. 

\textbf{Programming Problem Construction.} We adapted 50 multi-API problems from PanNumEval \citep{zan_diffcoder_nodate}, applying the same transformations to create semantically equivalent tasks using Monkey and BeatNum APIs. We carefully examine the transformation result and extend the transformation rules where some library information was not converted. Furthermore, we masked out the explicit API usage in task description like “… using np.linspace” to prevent unintentional hints about the original libraries.

\textbf{Execution Framework.} We implemented real-time bidirectional API transformation to enable execution-based evaluation using pseudo libraries, while maintaining the illusion of two new libraries for LLM's interactions. Specifically,
\begin{enumerate}
\item For executing the LLM generated code snippets: Pseudo API calls→Original API calls

\item For LLM obtaining the intermediate output\&exceptions: Original info $\rightarrow$ Pseudo info
\end{enumerate}

\subsubsection{Evaluation Results on MonkBeatEval}
We replicated representative related work that exhibited most competitive performance in Section \ref{subsec:RQ3_baselines} on MonkBeatEval using GPT-3.5. The results are presented in Table \ref{tab:MonkBeatEval-baselines}.

The trends observed in the results are consistent with the discussions in Section \ref{sec: experiments}, with ExploraCoder achieving SOTA performance.
This consistency further demonstrates the robustness of our evaluation, showcasing ExploraCoder's effectiveness and generalizability across different library contexts and integration scenarios.

\subsection{Evaluating prior API knowledge retention in API-pretrained and API-untrained base models}
\label{app: proof-of-knowledge-unseen}
We methodically differentiate between API-pretrained and API-untrained models based on publicly available information regarding their training data. However, to intuitively investigate whether the model retains substantial knowledge of Torchdata APIs, we directly ask the model to provide specific API details, as shown in Listing \ref{lst: API_know}. We set $temperature=0$ and $top\_p=1$ across the models. Our findings reveal that API-untrained models (GPT-3.5-turbo-0125 and GPT-4-0613) hallucinate API information, generating API names that do not exist in the Torchdata library. In contrast, the API-pretrained models (GPT-4-1106-preview, CodeQwen1.5-7B-Chat, deepseek-coder-6.7b-instruct) accurately generates correct API names, import paths, and even provides precise usage descriptions.
These results provide evidence supporting the validity of our experimental settings.
\begin{figure*}[htbp]
\begin{lstlisting}[caption={Demonstration of prior API knowledge in different LLMs.}, label={lst: API_know}]
what are some APIs in torchdata? Give me 5 specific API names(including full import path)
------------------------------------------------ Output by GPT-3.5-turbo-0125 ------------------------------------------------
1. torchdata.datasets.ImageFolderDataset
   - Full import path: torchdata.datasets.ImageFolderDataset
2. torchdata.datasets.CIFAR10Dataset
   - Full import path: torchdata.datasets.CIFAR10Dataset
3. torchdata.datasets.MNISTDataset
   - Full import path: torchdata.datasets.MNISTDataset
4. torchdata.transforms.Compose
   - Full import path: torchdata.transforms.Compose
5. torchdata.transforms.RandomHorizontalFlip
   - Full import path: torchdata.transforms.RandomHorizontalFlip
---------------------------------------------------- Output by GPT-4-0613 ----------------------------------------------------
1. torchdata.datasets.Files: This API is used to create a dataset from a list of files. It can be used to load and preprocess data from different file formats.
2. torchdata.datasets.Dataset: This is the base class for all datasets in torchdata. It provides methods for loading and iterating over data.
3. torchdata.datasets.Transform: This API is used to apply transformations to the data in a dataset. It can be used to preprocess data, such as normalizing or augmenting it.
4. torchdata.datasets.Folder: This API is used to create a dataset from a directory of files. It can be used to load and preprocess data from different file formats.
5. torchdata.datasets.ImageFolder: This API is used to create a dataset from a directory of images. It can be used to load and preprocess image data.
----------------------------------------------- Output by GPT-4-1106-preview -----------------------------------------------
1. `torchdata.datapipes.iter.FileOpener`:
   ```python
   from torchdata.datapipes.iter import FileOpener
   ```
   This API is used to open files from a given source (like a list of file paths) and yield a tuple of the file name and a file object.
2. `torchdata.datapipes.iter.HttpReader`:
   ```python
   from torchdata.datapipes.iter import HttpReader
   ```
   This API allows you to read data from HTTP URLs. It takes an iterable of URLs and yields the contents of the URLs.
3. `torchdata.datapipes.iter.CSVParser`:
   ```python
   from torchdata.datapipes.iter import CSVParser
   ```
   This API is used to parse CSV files. It takes an iterable of tuples (usually coming from `FileOpener`) and yields the parsed CSV content.
4. `torchdata.datapipes.iter.RoutedDecoder`:
   ```python
   from torchdata.datapipes.iter import RoutedDecoder
   ```
   This API is a decoder datapipe that applies different decoders based on the extension of the file path or a custom function provided by the user.
5. `torchdata.datapipes.iter.Batcher`:
   ```python
   from torchdata.datapipes.iter import Batcher
   ```
   This API batches data points into a specified batch size. It's useful for creating mini-batches of data for training machine learning models.
----------------------------------------------- Output by CodeQwen1.5-7B-Chat -----------------------------------------------
Here are five specific API names in torchdata, along with their full import paths:
1. `torchdata.datapipes.iter.IterDataPipe`: This is the base class for all iterable data pipes in torchdata.
2. `torchdata.datapipes.iter.Mapper`: This data pipe applies a function to each element in the input data pipe.
3. `torchdata.datapipes.iter.Filter`: This data pipe filters out elements from the input data pipe based on a condition.
4. `torchdata.datapipes.iter.Shuffler`: This data pipe shuffles the elements in the input data pipe.
5. `torchdata.datapipes.iter.Batcher`: This data pipe batches elements from the input data pipe into fixed-size groups.
------------------------------------------- Output by deepseek-coder-6.7b-instruct -------------------------------------------
TorchData is a package in PyTorch that provides data utilities and primitives to create complex data pipelines. Here are five specific APIs in torchdata:
1. `torchdata.datapipes.iter.FileLister`: This API is used to list files in a directory.
2. `torchdata.datapipes.iter.FileOpener`: This API is used to open files.
3. `torchdata.datapipes.iter.LineReader`: This API is used to read lines from a file.
4. `torchdata.datapipes.iter.CSVParser`: This API is used to parse CSV files.
5. `torchdata.datapipes.iter.MapDataPipe`: This API is used to apply a function to each element in the data pipe.
Please note that these are just a few examples. The full list of APIs in torchdata can be found in the [official documentation](https://pytorch.org/data/).
\end{lstlisting}
\end{figure*}

\subsection{Disccussion of fairness comparison between gpt-4-0613 and gpt-4-1106-preview.}
\label{app: gpt-4-compare-fairness}
GPT-4-0613 and GPT-4-1106-preview are two closely released version of GPT-4. According to publicly available information, the former is trained on data available up until September 2021, while the latter is a more recent version trained on data up until April 2023. In our experiment, we assume that both models share a similar architecture
, and that the performance gap of direct generation between the two is primarily due to the absence of API knowledge in training corpura, i.e. the performance gap between API-pretrianed and API-untrained models. Appendix \ref{app: proof-of-knowledge-unseen} has shown that while GPT-4-0613 is unaware of the Torchdata APIs, GPT-4-1106 can effectively recite the API details. In this context, we demonstrate in Section \ref{subsec: exp_RQ1} that integrating our ExploraCoder framework allows API-untrained models to surpass their API-pretrained counterparts, whereas integrating naive RAG does not, proving the effectiveness of ExploraCoder.

\subsection{Additional implementation details}
\label{app: implementation_details}
Torchdata is a library that facilitate multiple data processing operations. For task planning module, we ask GPT-3.5-turbo-0125 (API-untrained model) to summarize Torchdata's purpose, key concepts, and API division logic based on Torchdata's README page\footnote{\url{https://github.com/pytorch/data/blob/v0.7.1/README.md}}. The summarized results are presented in Listing \ref{lst: lib_summ}. We also extracted few-shot API invocation planners demonstrated in Listing \ref{lst: fewshot_planner} following \citet{ma_compositional_2024}'s approach. And both information are used for invocation task planning. Unlike the detailed functionalities for each APIs, the summarization and planners demonstrations give high-level insights into the library, facilitating better planning and reasoning for LLMs \citep{zheng_take_2023}. We use such summarization to represent limited domain knowledge for task planning, and no further detailed API usage information is leaked for problem solving. We also demonstrate ExploraCoder's prompts in Listing \ref{lst: prompt for subtask planner} - \ref{lst: prompt for final solution generator}.
\begin{lstlisting}[caption={Condensed introduction for Torchdata.}, label={lst: lib_summ}]
Torchdata is a library of common modular data loading primitives for constructing flexible data pipelines. It introduces composable Iterable-style and Map-style building blocks called DataPipes, which work well with PyTorch's DataLoader and have functionalities for loading, parsing, caching, transforming, and filtering datasets. 
DataPipes can be composed together into datasets and support execution in various settings and execution backends using DataLoader2. 
The library aims to make data loading components more flexible and reusable by providing a new DataLoader2 and modularizing features of the original DataLoader into DataPipes. 
DataPipes are a renaming and repurposing of the PyTorch Dataset for composed usage, allowing for easy chaining of transformations to reproduce sophisticated data pipelines. 
DataLoader2 is a light-weight DataLoader that decouples data-manipulation functionalities from torch.utils.data.DataLoader and offers additional features such as checkpointing/snapshotting and switching backend services for high-performant operations.
\end{lstlisting}
\begin{lstlisting}[caption={We demonstrate 2  examples for API invocation planner.}, label={lst: fewshot_planner}]
[task]
Read the contents of a file and verify its hash value.
[subtasks]
1. Open a file using FileOpener
2. Wrap the file object using IterableWrapper
3. Check the hash value of the file using check_hash
[task]
Fetch the first line of a text file from a given URL and print it alongside the URL.
[subtasks] 
1. Instantiate an OnlineReader datapipe using an IterableWrapper that holds the URL of the text file.
2. Read lines from the OnlineReader datapipe.
3. Iterate over the datapipe and output both the URL and the first line of the text file
\end{lstlisting}

\begin{lstlisting}[caption={prompt for subtask planner.}, label={lst: prompt for subtask planner}]
I will give you a task that needs interactions with external APIs. You need to break down the task into several subtasks that can be implemented by invoking APIs.
{library_summary}
Examples: {fewshot_examples}
Task: {Task}
Subtasks: 
\end{lstlisting}
\begin{lstlisting}[caption={prompt for CoAE.}, label={lst: prompt for CoAE}]
We have decomposed a user requirement into multiple subtasks and tested some api-calling codes for each subtask.
The user has prepared some external file you will need and defines the test inputs for you:
```
{example_inputs}
{code_context}\
```\
{prior_subtasks_exploration_experience}

Now you need to learn the API usage experience from previous subtasks and implement the subsequent subtask.
<subtask>{subtask_cnt}. {subtask}</subtask>

Here are some Torchdata APIs maybe useful:
{library_api_info}

Requirements:
1. Write a playground code that imports neccessary API(s), defines your own test data as input, and calls the APIs to implement the subtask. Wrap the code in a ```python block```.
2. For each used API, read the API description to learn the [data formats] and [semantics] of the input/output object. Make sure the object is converted to the correct format and semantics before passing it to an API.
3. Direclty use the user-defined example inputs as your playground code inputs. Make use of the explored APIs from prior subtasks and predefined functions for this subtask implementation.
4. You can print anywhere to check the the data or object format. Such output will be observed after execution.
\end{lstlisting}
\begin{lstlisting}[caption={prompt for CoAE self-debug.}, label={lst: prompt for CoAE self-debug}]
You were writing playground codes to explore external APIs usage for a subtask. Now you encountered an error. You need to debug the API usage and make the code executable.

## The buggy code:
```
{buggy_code}
```
## Error message:
{error_message}

## Relevant APIs
{api_list_str}

We omit the format requirement here.
\end{lstlisting}
\begin{lstlisting}[caption={prompt for final solution generator.}, label={lst: prompt for final solution generator}]
---------------------- system prompt ----------------------
# Context #
You are a senior Python programmer. You are assigned a task to implement an incomplete function to meet user's requirement. You find a new external library `Torchdata` in <<library_documents>> that is helpful.
To better learn the correct usage of Torchdata's APIs, you've thought of some relevant subtasks. For each <<subtask>>, you have first crafted a <<playground_code>> to call APIs to implement the subtask, then had an <<observation>> of the code's executability, execution output, and error message.
# Objective #
Now you need to implement the user <<requirement>> by importing neccessary APIs and completing the <<incomplete_function>>.
# Response #
Your response should contain a complete code snippet in the following format:
```python
[YOUR IMPORT HERE]
original incomplete code snippet
[YOUR COMPLETION HERE]
```
----------------------- user prompt ------------------------
You need to complete a function to meet requirement.
<requirement>
{requirement}
</requirement>
<incomplete_function>
{cg_task_prompt}
</incomplete_function>
You have explored some API usage on various subtasks:
<explorations_experience>
{subtask_exploration_list}
</explorations_experience>
Refer to relevant APIs information:
<library_documents>
{library_api_info}
</library_documents>
Now make use of the experience and supplemented APIs to complete the function.
Note that the subtasks may not directly related to the user requirement, excessive or unnecessary API calls may exist. But they are to help you understand the library's APIs behavior and usage. 
You have to reorganize API call sequence, add your own implementation to help transforming the data format between API calls.
\end{lstlisting}

\subsection{Case study}
\label{app: casestudy}
We have conducted a series of case studies, here we provide examples of different methods (naive RAG, ExploraCoder, Self-Repair) solving the same example problem from our benchmark. We also provided a case study of ExploraCoder*, where we demonstrate the self-debug trace at an failed intermediate subtask in CoAE. For each example, we provide discussion and analysis in the end of the listing.

\begin{lstlisting}[caption=A failed example for naive RAG. We omit the API signature and description for simplicity]
"""
Please complete the following function, here are some APIs maybe useful:
<API>
torchdata.datapipes.iter.ParagraphAggregator
torchdata.datapipes.map.Batcher
torchdata.datapipes.iter.Batcher
torchdata.datapipes.iter.OnDiskCacheHolder
torchdata.datapipes.iter.InBatchShuffler
torchdata.datapipes.iter.BucketBatcher
torchdata.datapipes.iter.JsonParser
torchdata.datapipes.iter.Shuffler
torchdata.datapipes.map.InMemoryCacheHolder
torchdata.datapipes.iter.FullSync
torchdata.datapipes.iter.CSVParser
torchdata.datapipes.iter.Prefetcher
torchdata.datapipes.iter.OnlineReader
torchdata.datapipes.iter.CSVDictParser
torchdata.datapipes.iter.ParquetDataFrameLoader
torchdata.datapipes.map.Shuffler
torchdata.datapipes.iter.RandomSplitter
torchdata.datapipes.map.Concater
torchdata.datapipes.iter.LineReader
torchdata.datapipes.iter.FileLister
</API>
"""
from torchdata.datapipes.iter import *
from typing import List, Tuple
import torch
def process_paragraphs_map_fn(paragraph: Tuple[str, str]) -> str:
    file_name = paragraph[0]
    paragraph_content = paragraph[1]
    return paragraph_content

def build_text_data_pipe(
        root: str="./torchdata-programming-tasks/task_82",  # Directory containing text files
        batch_size: int=16,  # Batch size
    ):
    """
    Load text data from files under a given directory, process the text data into paragraphs, create mini-batches of paragraphs, and shuffle the batches. The processed data should be cached and held in memory.
    """

Your response should contain a complete code snippet wrapped in```python```
------------------- GPT-3.5-turbo output -------------------
```python
# Define a function to process paragraphs
def process_paragraphs_map_fn(paragraph: Tuple[str, str]) -> str:
    file_name = paragraph[0]
    paragraph_content = paragraph[1]
    return paragraph_content

# Build the text data pipeline
def build_text_data_pipe(
        root: str="./torchdata-programming-tasks/task_82",  # Directory containing text files
        batch_size: int=16,  # Batch size
    ):
    # Create a DataPipe to list files
    file_list_dp = FileLister(root=root, masks=["*.txt"])
    # Create a DataPipe to read lines from files
    line_reader_dp = LineReader(file_list_dp)
    # Create a DataPipe to aggregate lines into paragraphs
    paragraph_aggregator_dp = ParagraphAggregator(line_reader_dp)
    # Create a DataPipe to batch paragraphs
    batcher_dp = Batcher(paragraph_aggregator_dp, batch_size=batch_size)
    # Create a DataPipe to shuffle batches
    shuffler_dp = InBatchShuffler(batcher_dp)
    # Create a DataPipe to cache and hold data in memory
    cached_dp = InMemoryCacheHolder(shuffler_dp)
    return cached_dp
------------------------- Analysis -------------------------
In this case, GPT-3.5-turbo + naive RAG successfully recalled relevant APIs to generator. However, it hallucinate on the usage of `FileLister` and `LineReader` APIs, since `FileLister` outputs the filename, while `LineReader` expects file stream as input to read each line from the stream. Such mistake hinders the successful generation of correct code solutions for multi-API problems.
\end{lstlisting}

\begin{lstlisting}[caption=A successful example for ExploraCoder. We show the formed exploration trace $\mathcal{\hat{E}}$ in CoAE\, and a successful final solution generated based on it.]
[requirement]
Load text data from files under a given directory, process the text data into paragraphs, create mini-batches of paragraphs, and shuffle the batches. The processed data should be cached and held in memory.
-------------------- Subtask 1 on CoAE --------------------
[subtask]
1. List all text files in the given directory
[exploratory code]
import torchdata
from torchdata.datapipes.iter import FileLister
root = "./torchdata-programming-tasks/task_82"
file_lister = FileLister(root=root)
for file_path in file_lister:
    print(file_path)
[observation]
Status: Executable
Output:
./torchdata-programming-tasks/task_82/file_0.txt
...omit 3 lines...
./torchdata-programming-tasks/task_82/file_4.txt
-------------------- Subtask 2 on CoAE --------------------
[subtask]
2. Open each text file and read the text data
[exploratory code]
import torchdata
from torchdata.datapipes.iter import FileLister, IoPathFileOpener, LineReader
root = "./torchdata-programming-tasks/task_82"
file_lister = FileLister(root=root)
file_opener = IoPathFileOpener(source_datapipe=file_lister)
line_reader = LineReader(source_datapipe=file_opener)
for file_name, line in line_reader:
    print(f"File: {file_name}")
    print(f"Content: {line}")
[observation]
Status: Executable
Output:
File: ./torchdata-programming-tasks/task_82/file_0.txt
Content: This is line 0 of file 0
File: ./torchdata-programming-tasks/task_82/file_0.txt
...omit 995 lines...
File: ./torchdata-programming-tasks/task_82/file_4.txt
Content: This is line 99 of file 4
-------------------- Subtask 3 on CoAE --------------------
[subtask]
3. Process the text data into paragraphs
[exploratory code]
import torchdata
from torchdata.datapipes.iter import FileLister, IoPathFileOpener, LineReader, ParagraphAggregator
from typing import Tuple
root = "./torchdata-programming-tasks/task_82"
file_lister = FileLister(root=root)
file_opener = IoPathFileOpener(source_datapipe=file_lister)
line_reader = LineReader(source_datapipe=file_opener)
paragraph_aggregator = ParagraphAggregator(source_datapipe=line_reader)
for file_name, paragraph in paragraph_aggregator:
    print(f"File: {file_name}")
    print(f"Paragraph: {paragraph}")
[observation]
Status: Executable
Output:
File: ./torchdata-programming-tasks/task_82/file_0.txt
Paragraph: This is line 0 of file 0
This is line 1 of file 0
...omit 500 lines...
This is line 98 of file 4
This is line 99 of file 4
-------------------- Subtask 4 on CoAE --------------------
[subtask]
4. Create mini-batches of paragraphs using torchdata APIs
[exploratory code]
import torchdata
from torchdata.datapipes.iter import FileLister, IoPathFileOpener, LineReader, ParagraphAggregator, Batcher
from typing import Tuple
root = "./torchdata-programming-tasks/task_82"
file_lister = FileLister(root=root)
file_opener = IoPathFileOpener(source_datapipe=file_lister)
line_reader = LineReader(source_datapipe=file_opener)
paragraph_aggregator = ParagraphAggregator(source_datapipe=line_reader)
batcher = Batcher(datapipe=paragraph_aggregator, batch_size=16)
for batch in batcher:
    print(batch)
[observation]
Status: Executable
Output:
[('./torchdata-programming-tasks/task_82/file_0.tx...is line 98 of file 4\nThis is line 99 of file 4')]
-------------------- Subtask 5 on CoAE --------------------
[subtask]
5. Shuffle the batches using shuffle API
[exploratory code]
import torchdata
from torchdata.datapipes.iter import FileLister, IoPathFileOpener, LineReader, ParagraphAggregator, Batcher, InBatchShuffler
root = "./torchdata-programming-tasks/task_82"
file_lister = FileLister(root=root)
file_opener = IoPathFileOpener(source_datapipe=file_lister)
line_reader = LineReader(source_datapipe=file_opener)
paragraph_aggregator = ParagraphAggregator(source_datapipe=line_reader)
batcher = Batcher(datapipe=paragraph_aggregator, batch_size=16)
shuffler = InBatchShuffler(datapipe=batcher)
for shuffled_batch in shuffler:
    print(shuffled_batch)
[observation]
Status: Executable
Output:
[('./torchdata-programming-tasks/task_82/file_4.tx...is line 98 of file 2\nThis is line 99 of file 2')]
--------------------- Subtask 6 on CoAE --------------------
[subtask]
6. Cache and hold the processed data in memory
[exploratory code]
import torchdata
from torchdata.datapipes.iter import FileLister, IoPathFileOpener, LineReader, ParagraphAggregator, Batcher, InMemoryCacheHolder
from typing import Tuple
root = "./torchdata-programming-tasks/task_82"
file_lister = FileLister(root=root)
file_opener = IoPathFileOpener(source_datapipe=file_lister)
line_reader = LineReader(source_datapipe=file_opener)
paragraph_aggregator = ParagraphAggregator(source_datapipe=line_reader)
batcher = Batcher(datapipe=paragraph_aggregator, batch_size=16)
cache_holder = InMemoryCacheHolder(source_dp=batcher)
for data in cache_holder:
    print(data)
[observation]
Status: Executable
Output:
[('./torchdata-programming-tasks/task_82/file_0.tx...is line 98 of file 4\nThis is line 99 of file 4')]
</explorations_experience>
Refer to relevant APIs information:
<library_documents>
{# We omit the API information for simplicity}
</library_documents>
Now make use of the experience and supplemented APIs to complete the function.
Note that the subtasks may not directly related to the user requirement, excessive or unnecessary API calls may exist. But they are to help you understand the library's APIs behavior and usage. You have to reorganize API call sequence, add your own implementation to help transforming the data format between API calls.
---------------------- Final Solution ----------------------
from torchdata.datapipes.iter import *
from typing import List, Tuple
import torch
def build_text_data_pipe(
        root: str="./torchdata-programming-tasks/task_82",  # Directory containing text files
        batch_size: int=16,  # Batch size
    ):
    file_lister = FileLister(root=root)
    file_opener = IoPathFileOpener(source_datapipe=file_lister)
    line_reader = LineReader(source_datapipe=file_opener)
    paragraph_aggregator = ParagraphAggregator(source_datapipe=line_reader)
    batcher = Batcher(datapipe=paragraph_aggregator, batch_size=batch_size)
    shuffler = InBatchShuffler(datapipe=batcher)
    cache_holder = InMemoryCacheHolder(source_dp=shuffler)
    return cache_holder
------------------------- Analysis -------------------------
We can find that during CoAE, ExploraCoder increamentally reuse APIs invocations from prior subtask, and experiments with new API invocations and observe their behaviour. Eventually, ExploraCoder manage to find a successful API exploration trace, helping it to generate correct final solution.
\end{lstlisting}

\begin{lstlisting}[caption=A successful example for intermediate self-debug on ExploraCoder.]
[API invocation subtask]
Load the text data from the zip file using load_text_data_from_zip
----------------- failed experimental code -----------------
[code]
import torchdata
from torchdata.datapipes.iter import FileLister, ZipArchiveLoader, FileOpener
from torchdata.datapipes.utils.common import StreamWrapper

# define test inputs
root = './torchdata-programming-tasks/task_89'
file_path = './torchdata-programming-tasks/task_89/file_0.zip'

# Load text data from the zip file
file_stream = open(file_path, 'rb')
zip_loader = ZipArchiveLoader([(file_path, StreamWrapper(file_stream))])
file_opener = FileOpener(zip_loader)

# Iterate over the file opener to get text data
for file_info in file_opener:
    file_path, file_stream = file_info
    text_data = file_stream.read()
    print(f"Text Data from {file_path}: {text_data}")

file_stream.close()
[onservation]
Status: failed
Output:
Error Message:
TypeError: Expected string type for pathname, but got <class 'tuple'>
This exception is thrown by __iter__ of FileOpenerIterDataPipe(datapipe=
ZipArchiveLoaderIterDataPipe, encoding=None, length=-1, mode='r')
Error traceback at:
 - line 15 in <module>: for file_info in file_opener:
 - line 212 in get_file_binaries_from_pathnames: raise TypeError("Expected string type for pathname, but got {}"

------------------- fixed by self-debug -------------------
[code]
import torchdata
from torchdata.datapipes.iter import FileLister, ZipArchiveLoader, FileOpener
from torchdata.datapipes.utils.common import StreamWrapper

# User-defined test inputs
root = './torchdata-programming-tasks/task_89'
file_path = './torchdata-programming-tasks/task_89/file_0.zip'

# Load text data from the zip file
file_stream = open(file_path, 'rb')
zip_loader = ZipArchiveLoader([(file_path, StreamWrapper(file_stream))])

# Iterate over the zip loader to get text data
for file_info in zip_loader:
    file_path, file_stream = file_info
    text_data = file_stream.read()
    print(f"Text Data from {file_path}: {text_data}")
[observation]
Status: success
Output:
Text Data from torchdata-programming-tasks/task_89/
file_0.zip/file_0_0.txt: b'...'
...omit 3 lines...
tasks/task_89/file_0.zip/file_0_4.txt: b'...'
------------------------- Analysis -------------------------
In ExploraCoder's initial experimental code, it incorrectly hallucinates the usage of FileOpener. However, after a round of self-debugging, ExploraCoder is able to correct this simple API misuse and successfully observe behavior from the correct API invocation.
\end{lstlisting}

\begin{lstlisting}[
    caption=A failed example for Self-Repair.,
    firstline=1,
    lastline=98
]
[requirement]
Load text data from files under a given directory, process the text data into paragraphs, create mini-batches of paragraphs, and shuffle the batches. The processed data should be cached and held in memory.
-------------------- initial candidate --------------------
[code]
from torchdata.datapipes.iter import *
from typing import List, Tuple
import torch
def process_paragraphs_map_fn(paragraph: Tuple[str, str]) -> str:
    file_name = paragraph[0]
    paragraph_content = paragraph[1]
    return paragraph_content
def build_text_data_pipe(
        root: str="./torchdata-programming-tasks/task_82",  # Directory containing text files
        batch_size: int=16,  # Batch size
    ):
    data_pipe = FileLister(root) >> IoPathFileLister() >> FileOpener() >> LineReader()
    data_pipe = data_pipe.map(process_paragraphs_map_fn) >> Batcher(batch_size=batch_size) >> InMemoryCacheHolder()
    data_pipe = data_pipe >> Shuffler()
    return data_pipe
list_dp = list(build_text_data_pipe(root="./torchdata-programming
-tasks/task_82", batch_size=4))
[execution result]
Status: failed
Error message: 
TypeError: __init__() missing 1 required positional argument: 'root'
Error traceback at:
 - line 23 in <module>: list_dp = list(build_text_data_pipe(root="./torchdata-programming
 -tasks/task_82", batch_size=4))
 - line 18 in build_text_data_pipe:     data_pipe = FileLister(root) >> IoPathFileLister() >> FileOpener() >> LineReader()
------------------- round 1 self-repair -------------------
[feedback] 
The error message indicates that the `FileLister` class in the `build_text_data_pipe` function is missing the required positional argument `root`. The `FileLister` class expects the root directory containing text files as an argument. To resolve this error, make sure to provide the `root` argument when creating the `FileLister` instance in the `build_text_data_pipe` function. 
[code]
from torchdata.datapipes.iter import *
from typing import List, Tuple
import torch

def process_paragraphs_map_fn(paragraph: Tuple[str, str]) -> str:
    file_name = paragraph[0]
    paragraph_content = paragraph[1]
    return paragraph_content

def build_text_data_pipe(
        root: str="./torchdata-programming-tasks/task_82",  # Directory containing text files
        batch_size: int=16,  # Batch size
    ):
    """
    Load text data from files under a given directory, process the text data into paragraphs, create mini-batches of paragraphs, and shuffle the batches. The processed data should be cached and held in memory.
    """
    data_pipe = FileLister(root=root) >> IoPathFileLister() >> FileOpener() >> LineReader()
    data_pipe = data_pipe.map(process_paragraphs_map_fn) >> Batcher(batch_size=batch_size) >> InMemoryCacheHolder()
    data_pipe = data_pipe >> Shuffler()
    
    return data_pipe
list_dp = list(build_text_data_pipe
(root="./torchdata-programming-tasks/task_82", batch_size=4))
Status: failed
Error message: 
TypeError: __init__() missing 1 required positional argument: 'root'
Error traceback at:
 - line 23 in <module>: list_dp = list
 (build_text_data_pipe(root="./torchdata-programming-
 tasks/task_82", batch_size=4))
 - line 18 in build_text_data_pipe:     data_pipe = FileLister(root=root) >> IoPathFileLister() >> FileOpener() >> LineReader()
------------------- round 2 self-repair -------------------
[feedback]
The error message indicates that the `FileLister` class in the `build_text_data_pipe` function is missing the required positional argument `root`. The `FileLister` class expects the root directory containing text files as an argument. To resolve this error, make sure to provide the `root` argument when creating the `FileLister` instance in the `build_text_data_pipe` function. 
[code]
from torchdata.datapipes.iter import *
from typing import List, Tuple
import torch
def process_paragraphs_map_fn(paragraph: Tuple[str, str]) -> str:
    file_name = paragraph[0]
    paragraph_content = paragraph[1]
    return paragraph_content
def build_text_data_pipe(
        root: str="./torchdata-programming-tasks/task_82",  # Directory containing text files
        batch_size: int=16,  # Batch size
    ):
    """
    Load text data from files under a given directory, process the text data into paragraphs, create mini-batches of paragraphs, and shuffle the batches. The processed data should be cached and held in memory.
    """
    data_pipe = FileLister(root=root) >> IoPathFileLister() >> FileOpener() >> LineReader()
    data_pipe = data_pipe.map(process_paragraphs_map_fn) >> Batcher(batch_size=batch_size) >> InMemoryCacheHolder()
    data_pipe = data_pipe >> Shuffler()
    return data_pipe
list_dp = list(build_text_data_pipe(root="./torchdata
-programming-tasks/task_82", batch_size=4))
[execution result]
Status: failed
Error message:
TypeError: __init__() missing 1 required positional argument: 'root'
Error traceback at:
 - line 23 in <module>: list_dp = list(build_text_data_pipe(root="./torchdata-programming
 -tasks/task_82", batch_size=4))
 - line 18 in build_text_data_pipe:     data_pipe = FileLister(root=root) >> IoPathFileLister() >> FileOpener() >> LineReader()
------------------- round 3 self-repair -------------------
[feedback]
\end{lstlisting}
\begin{lstlisting}[caption=A failed example for Self-Repair (Cont.).]
The error message indicates that the `build_text_data_pipe` function is missing the required positional argument `root`. In this case, the `FileLister` class expects the `root` argument to be provided. To resolve this issue, make sure to pass the `root` argument when calling the `build_text_data_pipe` function. 
[code]
from torchdata.datapipes.iter import *
from typing import List, Tuple
import torch
def process_paragraphs_map_fn(paragraph: Tuple[str, str]) -> str:
    file_name = paragraph[0]
    paragraph_content = paragraph[1]
    return paragraph_content
def build_text_data_pipe(
        root: str = "./torchdata-programming-tasks/task_82",  # Directory containing text files
        batch_size: int = 16,  # Batch size
    ):
    """
    Load text data from files under a given directory, process the text data into paragraphs, create mini-batches of paragraphs, and shuffle the batches. The processed data should be cached and held in memory.
    """
    data_pipe = FileLister(root=root) >> IoPathFileLister() >> FileOpener() >> LineReader()
    data_pipe = data_pipe.map(process_paragraphs_map_fn) >> Batcher(batch_size=batch_size) >> InMemoryCacheHolder()
    data_pipe = data_pipe >> Shuffler()
    return data_pipe
list_dp = list(build_text_data_pipe(
root="./torchdata-programming-tasks/task_82", batch_size=4))
------------------ round 4-5 self-repair ------------------
...omit for simplicity...
------------------- round 6 self-repair -------------------
[feedback]
The error message indicates that the 'root' argument is missing when creating the FileLister data pipe. The FileLister API expects the root argument to be provided, but it is not being passed correctly in the build_text_data_pipe function. To resolve this issue, ensure that the 'root' argument is correctly passed when creating the FileLister data pipe within the build_text_data_pipe function. 
[code]
from torchdata.datapipes.iter import *
from typing import List, Tuple
import torch

def process_paragraphs_map_fn(paragraph: Tuple[str, str]) -> str:
    file_name = paragraph[0]
    paragraph_content = paragraph[1]
    return paragraph_content

def build_text_data_pipe(
        root: str = "./torchdata-programming-tasks/task_82",  # Directory containing text files
        batch_size: int = 16,  # Batch size
    ):
    """
    Load text data from files under a given directory, process the text data into paragraphs, create mini-batches of paragraphs, and shuffle the batches. The processed data should be cached and held in memory.
    """
    data_pipe = FileLister(root=[root]) >> IoPathFileLister() >> FileOpener() >> LineReader()
    data_pipe = data_pipe.map(process_paragraphs_map_fn) >> Batcher(batch_size=batch_size) >> InMemoryCacheHolder()
    data_pipe = data_pipe >> Shuffler()
    return data_pipe
list_dp = list(build_text_data_pipe(
root="./torchdata-programming-tasks/task_82", batch_size=4))
[execution result]
Status: failed
Error message: 
TypeError: __init__() missing 1 required positional argument: 'root'
Error traceback at:
 - line 23 in <module>: list_dp = list(build_text_data_pipe(
 root="./torchdata-programming-tasks/task_82", batch_size=4))
 - line 18 in build_text_data_pipe:     data_pipe = FileLister(root=[root]) >> IoPathFileLister() >> FileOpener() >> LineReader()
------------------------- Analysis -------------------------
In the initial code solution, Self-Repair severely misuses the Torchdata library. Specifically, it incorrectly assumes that Torchdata APIs can be chained using “>>”, and this incorrect API usage pattern propagates throughout the entire LLM-decoded API invocation sequence. Despite receiving clear error messages in each round of Self-Repair, the system fails to correct the code because its initial solution deviates significantly from the canonical implementation.
\end{lstlisting}

\end{document}